\documentclass[prd,nofootinbib,floats,superscriptaddress,eqsecnum,tightenlines,preprintnumbers,11pt]{revtex4-1}

\usepackage{mathtools}
\usepackage{amssymb}
\usepackage{dsfont}
\usepackage{lipsum}
\usepackage{mathtools}
\usepackage{xcolor}
\usepackage[colorlinks=true,linkcolor=blue,citecolor=magenta,linktocpage=true]{hyperref}
\usepackage{physics}
\usepackage{csquotes}
\usepackage{mathtools, stmaryrd}
\usepackage{bm}
\usepackage{array}
\usepackage{multirow}
\usepackage{tensor}

\newcolumntype{M}[1]{>{\centering\arraybackslash$\displaystyle}p{#1}<{$}}

\usepackage{caption}
\usepackage{subcaption}

\usepackage[makeroom]{cancel}

\def\be{\begin{equation}}
	\def\ee{\end{equation}}
\def\bea{\begin{eqnarray}}
	\def\eea{\end{eqnarray}}
\def\beq{\begin{eqnarray}}
	\def\eeq{\end{eqnarray}}
\def\bas{\begin{subequations}\begin{eqnarray}}
		\def\eas{\end{eqnarray}\end{subequations}}
\def\nn{\nonumber}
\def\rd{\mathrm{d}}
\def\veps{\varepsilon}

\newcommand{\A}{{\mathcal A}}

\newcommand{\cE}{{\mathcal E}}

\newcommand{\cI}{{\mathcal I}}

\newcommand{\cM}{{\mathcal M}}

\newcommand{\po}[1]{{#1}_{\mathrm{p}}}
\newcommand{\ax}[1]{{#1}_{\mathrm{a}}}

\newcommand{\cO}{{\mathcal O}}

\newcommand{\D}{\partial}
\newcommand{\Db}{\bar{\partial}}

\def\nn{\nonumber}

\begin{document}

\title{Quadratic perturbations of the Schwarzschild black hole: \\ \smallskip The algebraically special sector}

	\author{Jibril Ben Achour}
	\affiliation{Arnold Sommerfeld Center for Theoretical Physics, Munich, Germany}
	\affiliation{Munich Center for Quantum Science and Technology, Munich, Germany}
	\affiliation{Univ de Lyon, ENS de Lyon, Laboratoire de Physique, CNRS UMR 5672, Lyon 69007, France}
	\author{Hugo Roussille}
	\affiliation{Univ de Lyon, ENS de Lyon, Laboratoire de Physique, CNRS UMR 5672, Lyon 69007, France}


	\begin{abstract}
We investigate quadratic algebraically special perturbations (ASPs) of the Schwarzschild black hole. Their dynamics are derived from the expansion up to second order in perturbation of the most general algebraically special twisting vacuum solution of general relativity. Following this strategy, we present analytical expressions for the axial-axial, polar-polar and polar-axial source terms entering in the dynamical equations. We show that these complicated inhomogeneous equations can be solved analytically and we present explicit expressions for the profiles of the quadratic ASPs. As expected, they exhibit exponential growth both at the past and future horizons even in the non-linear regime. We further use this result to analyze the quadratic zero modes and their interpretation in terms of quadratic corrections to mass and spin of the Schwarzschild black hole. The present work provides a direct extension beyond the linear regime of the original work by Couch and Newman. 
	\end{abstract}

\maketitle
\tableofcontents

\newpage

\section*{Introduction}

The gravitational radiation emitted during the coalescence of a binary black hole merger is well described by black hole perturbation theory at both early and late time,  corresponding respectively to the inspiral phase and the relaxation to equilibrium. For spherically symmetric black holes, the perturbation theory was set up in the late fifties by Regge and Wheeler in their seminal paper \cite{Regge:1957td} where the dynamics of the linearized perturbations around the Schwarzschild black hole were described. It is now well understood that during the relaxation to equilibrium, the perturbed Schwarzschild black hole radiates away all its higher multipoles \cite{Martel:2005ir}. While the early time signal does depend on the details of the merger, the intermediate signal is dominated by damped waves dubbed quasi-normal modes (QNMs) followed by power-law tails which drive the late-time signal. The QNMs frequencies are universal and fully characterized by the black hole mass (and spin in the case of a rotating black hole) which makes the QNM spectrum a unique footprint of the black hole geometry. Therefore the potential observation of long-lived QNMs in the near future provides a powerful test of general relativity and its possible extensions \cite{Berti:2004md, Berti:2009kk, Berti:2015itd, Franchini:2023eda}. 

While the description of the relaxation process after the merger based on the linear regime of the perturbation theory is well developed, several points are still debated such as a characterization of the time domain during which QNMs actually dominate the signal or their stability against small deformations \cite{Cardoso:2024mrw, Destounis:2021lum, Berti:2022xfj, Oshita:2024fzf}. More recently, characterizing non-linear corrections to the QNMs spectrum during the relaxation process right after the merger has attracted much attention\footnote{See \cite{DeLuca:2023mio, Riva:2023rcm} for related investigations in the static regime.}. On the one hand, this requires pushing black hole perturbation theory up to second order, a task which was undertaken by various authors in the last decades \cite{Gleiser:1995gx, Gleiser:1998rw, Campanelli:1998jv, Garat:1999vr,  Nicasio:2000ge, Thompson:2016fxe, Brizuela:2007zza, Brizuela:2009qd}. On the other hand, the non-linearities can be investigated numerically, a target which has attracted considerable attention recently \cite{London:2014cma, Cheung:2022rbm, Mitman:2022qdl, Ma:2022wpv, Lagos:2022otp, Bhagwat:2023fid, Kehagias:2023ctr, Khera:2023oyf, Redondo-Yuste:2023seq, Perrone:2023jzq, Bucciotti:2023ets, Cheung:2023vki, Qiu:2023lwo, Zhu:2024rej, Ma:2024qcv, Zhu:2024dyl, Rosato:2024arw} (see \cite{Yi:2024elj} for a recent prospect on the detectability of these non-linearities). While the spectrum of quadratic QNMs (dubbed QQNMs) has a straightforward relationship to the linear spectrum, the fate of the amplitude of quadratic perturbations has been addressed only recently \cite{Bucciotti:2024zyp, Bourg:2024jme}. Numerical investigations have shown that the amplitude of the quadratic modes depends on the multipole structure, dominating in some cases the one of their linear counterparts for large multipoles~\cite{London:2014cma, Cheung:2022rbm, Mitman:2022qdl, Ma:2022wpv}. This growing body of results reveals that the black hole spectroscopy program is more delicate than anticipated and underlines the necessity to further investigate the response of black holes to perturbations beyond the linear regime. 

In this work, we shall focus on the characterization of a certain class of quadratic perturbations which play a special role in black hole perturbation theory. It corresponds to the perturbations which, while spoiling the Petrov type D character of the background, preserve its algebraically special character allowing for a perturbed black hole geometry of Petrov type II\footnote{Generic perturbations usually break the algebraically special character of the background geometry such that no symmetry survives \cite{Araneda:2015gsa}}. For this reason, these perturbations are called \textit{algebraically special perturbations} (ASPs). So far, these modes have been discussed for the Schwarzschild black hole only at the linearized level first by Chandrasekhar in \cite{Chandrasekhar1984}, and later on by Couch and Newman in \cite{Couch:1973zc} (see also \cite{Qi:1993ey} for a more recent discussion). The case of the linear ASPs of the Kerr black hole was investigated by Wald in \cite{Wald:1973wwa}. The first goal of this work is to investigate the Schwarzschild ASPs at the quadratic level. Studying these modes beyond the linear regime can be motivated from several corners.

First, they stand aside from the other perturbations as they can be derived from pure symmetry arguments, without invoking any boundary conditions. As we shall see, they can be obtained by expanding to a given order in perturbation theory the most general algebraically special twisting vacuum solutions of general relativity (thus being of Petrov type II). Therefore, contrary to the other perturbations, one has already access to a fully non-perturbative description of these modes. This allows one to investigate how the non-linearities manifest in the perturbative treatment, providing a rather unique window to address this question.

Second, in dimension greater than 4, it has been shown that ASPs are purely stationary, i.e. they stand as pure zero modes connecting different higher dimensional black hole solutions \cite{Dias:2013hn}.  However, in the four dimensional case, they correspond to propagating, albeit non-oscillating modes. This rather peculiar property which shows up only in 4d comes with another feature. At the linear level, these ASPs diverge either on the future or on the past Schwarzschild horizon. This divergence prevent them from being well defined excitations of the Schwarzschild black hole which allow for a relaxation to equilibrium after a certain time. As such, they stand outside the standard perturbations of compact support one usually considers in black hole perturbation theory. For this reason, they are generally considered as pathological\footnote{Notice that while the standard quasi-normal modes also diverge on the past and future horizons, this divergence is non-physical in the sense that the QNMs only represent part of the full signal. The full perturbation is actually well-behaved as it corresponds to a perturbation with a compact support. When all the contributions are taken into account, the time-domain signal does not have any divergence by construction. On the contrary, the ASPs discussed here are exact analytic solutions and represent the full signal. Therefore, their divergence is pathological while the one of the standard QNMs is only artificial.}. At best, one can consider these perturbations as valid only on a finite portion of the null cone, as initially argued in \cite{Couch:1973zc}. 

Nevertheless, it turns out that among the well-defined perturbations of the Schwarzschild black hole, one can find specific excitations with similar spectral properties, namely a purely imaginary frequency which is numerically very close from the theoretical damping time of the Schwarzschild ASP \cite{MaassenvandenBrink:2000iwh}. Whether the value of the damping time matches exactly the ASP one is a matter of debate, but the main point is that one can exhibit well-defined highly damped modes of the Schwarzschild black hole which are spectrally close to the ASP \cite{Leung:1999fr, Leung:2003eq, MaassenvandenBrink:2000iwh} (see \cite{Berti:2003jh, Cook:2016ngj, Cook:2016fge, Hod:2013fea} for the Kerr black hole). Therefore, in view of the growing interest in the non-linear effects on the QNM spectrum, it is an interesting task to describe the behavior of the quadratic ASPs of the Schwarzschild black hole to later confront them to the QQNMs spectrum. This is one of the main goals of this work. Let us further stress that these algebraically special modes also play a key role in the understanding of the symmetries of black hole perturbations  and in particular in the proof of the isospectrality of the QNM spectrum \cite{ISO, Glampedakis:2017rar, Yurov:2018ynn, Lenzi:2021njy, Lenzi:2023inn}. Hence, understanding their non-linear behavior could also shed light on possible hidden structures of the QQNMs spectrum.


The paper is organized as follows. We begin in Section~\ref{sec1} by a brief review of the algebraically special modes as solutions of the RW and Zerilli wave equations. Then, we present in Section~\ref{sec2} the most general algebraically special twisted vacuum solutions of general relativity which will be our starting point to derive the profile of the quadratic ASPs. Section~\ref{sec3} presents the concrete dynamics of the linear ASPs, reviewing the results of Couch and Newman \cite{Couch:1973zc}. We use this opportunity to present in section~\ref{sec4} a concise and clear derivation of the action of the monopole and dipole linear zero modes. Then, in Section~\ref{sec5} we derive the dynamical equations for the quadratic ASPs and present the explicit analytical solutions for their profiles. We conclude this section by a careful treatment of the quadratic zero modes and we use this result to discuss the non-linear behavior of the dynamical quadratic ASPs. Section~\ref{sec6} contains a discussion of our results and the perspectives they open. Appendix~\ref{app:Ylm-stereo} is dedicated to a review of the main properties of real spherical harmonics. Finally, the reconstruction of the metric perturbations in the standard Regge-Wheeler parametrization \cite{Regge:1957td} is presented for completeness in appendix~\ref{app:exp-RW} and the detailed derivation of the explicit example is given in appendix~\ref{examp}.


\newpage
\section{Review of algebraically special modes}

\label{sec1}

Linear perturbations of the Schwarzschild black hole geometry are described by the well-known Regge-Wheeler and Zerilli equations. Using the spherical symmetry of the background, these two equations reduce to simple 2d second order differential equations dictating the time and radial profile of the polar  (even parity) perturbation, denoted $Y_{\text{RW}}$, and the axial (odd-parity) perturbations denoted $Y_{\text{Z}}$. These decoupled equations read respectively
\begin{align}
	&-\dv[2]{Y_\mathrm{RW}}{t} + \dv[2]{Y_\mathrm{RW}}{r_*} = V_\mathrm{RW}(r) Y_\mathrm{RW} \,,\label{eq:RW}\\
	&-\dv[2]{Y_\mathrm{Z}}{t} + \dv[2]{Y_\mathrm{Z}}{r_*} = V_\mathrm{Z}(r) Y_\mathrm{Z} \,,\label{eq:Z}
\end{align}
Here, $r_{\ast}$ is the tortoise coordinate defined by
\begin{equation}
	\dv{r_*}{r} = \Big(1 - \frac{2M}{r}\Big)^{-1} \qq{or} r_*(r) = r + 2M \log(r - 2M) \,,
\end{equation}
such that $r_{\ast} \rightarrow - \infty$ at the horizon, i.e. when $r=2M$ and the RW and Zerilli potentials are given explicitly by
\begin{align}
	V_\mathrm{RW}(r) &= \Big(1-\frac{2M}{r}\Big) \frac{2(\ax{\lambda} + 1)r - 6M}{r^3} \,,\\
	V_\mathrm{Z}(r) &= \Big(1-\frac{2M}{r}\Big) \frac{72M^3 + 72M^2 r \po{\lambda} + 24 M r^2 \po{\lambda}^2 + 8 r^3 \po{\lambda}^2 (\po{\lambda} + 1)}{r^3 (6M + 2 r \po{\lambda})^2} \,.
\end{align}
We have  written $\ax{\lambda}$ and $\po{\lambda}$ the values of $\lambda$ associated respectively to axial and polar perturbations.
In these expressions, we use the integer $\lambda$ defined from the angular momentum $\ell$ of perturbations by
\begin{equation}
	2\lambda = \ell(\ell+1) - 2 \,.
	\label{eq:def-lambda}
\end{equation}
Solving these two wave equations requires choosing some boundary conditions which lead to qualitatively different perturbations. The quasi-normal modes (QNM) are obtained by demanding that the wave is purely in-going at the horizon and purely out-going at $\cI^{+}$. Another choice consists in imposing that the wave is purely in-going at the horizon and at $\cI^{+}$ which gives the so called totally transmitted from the right modes (TTM$_R$). Demanding instead that the wave is purely out-going at both boundaries gives the so called totally transmitted from the left modes (TTM$_L$). 

Another way to solve for these equations is to introduce the ansätze
\begin{equation}
\label{ASP}
	Y_\mathrm{RW} = Y_\mathrm{RW}^\mathrm{ret} = e^{\ax{\kappa} (t - r_*)} \ax{f}(r) \qq{and} Y_\mathrm{Z} = Y_\mathrm{Z}^\mathrm{ret} = e^{-\po{\kappa} (t - r_*)} \po{f}(r) \,.
\end{equation}
where $\ax{\kappa}$ and $\po{\kappa}$ are real constants.
Injecting these expressions into \eqref{eq:RW} and~\eqref{eq:Z}, one obtains the following exact solutions
\begin{equation}
	\kappa = \frac{(\ell - 1)\ell(\ell + 1)(\ell + 2)}{12 M} \,,\quad \ax{f} = \frac{3M + \ax{\lambda} r}{r} \qq{and} \po{f} = \frac{r}{3M + \po{\lambda} r} \,.
\end{equation}
These solutions are dubbed \emph{algebraically special modes} for reasons that will be clarified in the following. They correspond to non-oscillating modes which violate the boundary conditions satisfied by the QNMs and by the TTMs$_{L/R}$, giving rise to a new family of modes. Indeed, introducing the retarded and advanced time coordinates 
\be
u = t - r_{\ast} \qquad v = t + r_{\ast}
\ee
we see from (\ref{ASP}) that while $Y_\mathrm{RW}$ diverges at $\cI^{+}$, i.e. when $u \rightarrow +\infty$, the polar perturbation $Y_\mathrm{Z}$ diverges at the horizon, i.e. when $u \rightarrow - \infty$. This couple of ASM comes with another set of solutions obtained by reversing the time coordinate $t$, leading to  
\begin{equation}
	Y_\mathrm{RW} = Y_\mathrm{RW}^\mathrm{adv} = e^{- \ax{\kappa} (t + r_*)} \ax{f}(r) \qq{and} Y_\mathrm{Z} = Y_\mathrm{Z}^\mathrm{adv} = e^{\po{\kappa} (t + r_*)} \po{f}(r) \,.
\end{equation}
This second couple of solutions behaves exactly in the opposite way from the former.

The status of these ASM, which are well defined exact solutions of the RW and Zerilli equations (at least when considered in a finite region of spacetime corresponding to a finite range of $u$ or $v$)  has attracted considerable attention over the years. Because they blow up on the horizon, these solutions of the RW and Zerilli equations cannot be considered as perturbations per se, as their evolution prevent the geometry to return back to the Schwarzschild one. Nevertheless, it is known that among the well behaved perturbations of the Schwarzschild black hole, one can find QNMs and TTMs exhibiting a vanishing frequency and a decay time which is numerically close from the ASM \cite{Leung:1999fr, MaassenvandenBrink:2000iwh, Leung:2003eq}, giving rise to well-behaved highly damped modes (see \cite{Cook:2016ngj, Cook:2016fge, Hod:2013fea} for Kerr). Despite these results, the exact spectral equivalence of these specific modes to the ASM remains an open question \cite{Berti:2009kk}.

More importantly for the purpose of this work, this set of ASM turn out to correspond to linearized algebraically special radiative solutions of general relativity, hence their name. 
This result was shown by Couch and Newman in \cite{Couch:1973zc} and their relation to the Chandrasekhar algebraically special modes was stressed by Qi and Schultz in \cite{Qi:1993ey}. This alternative derivation of these ASM makes it explicit that contrary to QNM and TTMs, one can derive their expressions without imposing boundary conditions, but simply by imposing symmetry restriction (i.e. restricting the Petrov type of the perturbed geometry). Remarkably, this strategy can be generalized to higher order in perturbation theory, giving access to an analytic treatment of the non-linear ASMs. As we shall see, studying these ASM beyond the linear order allows one to contemplate the complicated mode-mixing in an analytical form as well as to discuss the fate of non-linear zero modes in black hole perturbation theory.


\section{Twisting radiative spacetime and black hole perturbations}

\label{sec2}

In this section, we present the exact family of twisting vacuum radiative solutions of general relativity which will be our starting point to derive analytic expression for the quadratic algebraically special perturbations (ASP). We follow closely \cite{Stephani:2003tm} where this family of solutions is presented in detail. Let us introduce the tetrad $\omega $
given by
\begin{equation}
	\omega^l = \dd{u} + L \dd{z} + \bar{L} \dd{\bar{z}} \,, \quad \omega^n = \dd{r} + W \dd{z} + \bar{W} \dd{\bar{z}} + H \omega^l \,, \quad	\omega^m = \frac{\dd{\bar{z}}}{P \rho} \,,
\end{equation}
where the functions $L$, $W$, $H$, $P$ and $\rho$ defined by
\begin{align}
	\rho &= -\frac{1}{r + i\Sigma} \,, &2i\Sigma &= P^2 (\Db L - \D\bar{L}) \,,\nonumber\\
	W &= \frac{\partial_u L}{\rho} + \D(i\Sigma) \,,  &\D &= \partial_z - L \partial_u \,, \nonumber\\
	H &= \frac{K}{2} - r \partial_u(\log(P)) - \frac{M r + N \Sigma}{r^2 + \Sigma^2} \,, & K &= 2 P^2 \Re(\D(\Db\log P - \partial_u \bar{L})) \,,\nonumber\\
	N &= \Sigma K + P^2 \Re(\D\Db \Sigma - 2\partial_u\bar{L}\D\Sigma - \Sigma \partial_u\D\bar{L}) \,.
	\label{eq:def-funcs-metric}
\end{align}
Consider the line element given by
\begin{equation}
	\dd{s}^2 = -2 \omega^l \omega^n + 2 \omega^m \omega^{\bar{m}} \,,
	\label{eq:metric}
\end{equation}
In this setup, $M(u,z,\bar{z})$ and $P(u, z, \bar{z})$ are real functions and $L(u, z, \bar{z})$ is a complex function. Let us describe the main kinematical properties of the geometries belonging to this family. By construction, one has $\Psi_0 = \Psi_1 =0$ such that the geometry is algebraically special. The most general examples are of Petrov type II. The only non-vanishing Weyl scalars are given by 
\begin{align}
\Psi_2 & = (m + i N) \rho^3 \\
\Psi_3 & = - P^3 \rho^2 \partial I + \cO(\rho^3) \\
\Psi_4 & = P^2 \rho \partial_u I + \cO(\rho^2)
\end{align}
The Newman-Penrose spin coefficients (NPSC) satisfy
\begin{align}
\kappa = \sigma = \tau =0
\end{align}
while the other NPSC are non-vanishing. In particular, one has $\text{Im}(\rho) \neq 0$ which implies that the geometry is twisting.
It follows that the null tetrad describes a geodesic, non-shearing but twisting and expanding congruence of null rays. This family of solutions describes the twisting algebraically special vacuum radiative solutions of GR. When one further imposes that the twist vanishes, i.e, that $L=0$, this family of solutions reduces to the Robinson-Trautman geometries of Petrov type III.

One can show that under the above assumptions (choice of tetrad), the Einstein field equations reduce to
\begin{align}
	&i \D N - 3(M + iN) \partial_u L = 0 \,,\label{eq:evol-L}\\
	&P\qty[\D + 2(\D\log P - \partial_u\bar{L})] \D\qty[\Db(\Db\log P - \partial_u\bar{L}) + (\Db\log P - \partial_u\bar{L})^2] - \partial_u\qty[P^{-3}(M + i N)] = 0 \,,\label{eq:evol-P}
\end{align}
which provides the dynamical equations relating the three functions $(M,P,L)$. When $L=0$, the twist vanishes and this set of solutions reduces to the Robinson-Trautman family of exact radiative solutions in which case the mass $M$ can only depend on $u$.
How can this family of exact non-linear radiative solutions be useful to investigate black hole perturbations ?

When investigating black hole perturbations, one is interested in perturbations about a Petrov type D geometry (being either Kerr or Schwarzschild). Without further restriction, general perturbations will break the algebraically special character of the background geometry resulting in a Petrov type I geometry \cite{Araneda:2015gsa}. However, it is interesting to focus on the subset of these perturbations which, while spoiling the Petrov type D character of the background, still allows the perturbed background to remain algebraically special. Since the perturbed black hole geometrie remains at most of Petrov type II, the above family of exact solutions provides a framework to analyze these algebraically special perturbations without imposing any boundary conditions. 

This strategy was used by Couch and Newman in \cite{Couch:1973zc} to derive the linear ASP of the Schwarzschild black hole and soon after, the same approach was generalized to the linear ASP of the Kerr black hole by Wald \cite{Wald:1973wwa}. In the following, we shall focus on the Schwarzschild perturbations. For completeness, we will first review the analysis of Couch and Newman for the linear ASP. Then, we will generalize their result to second order in perturbation theory and present the exact solutions for the quadratic ASP.

\section{Linear algebraically special perturbations}

\label{sec3}

We start by reviewing the derivation of the linear ASP for the Schwarzschild background. Within this large family of twisting vacuum radiative solutions described by the two functions $(M, P,L)$, the Schwarzschild geometry corresponds
 \begin{equation}
P = P_0 = \frac{1}{\sqrt{2}} (1 + z\bar{z}) \qq{and} L = 0 \,.
\end{equation} 
while $M$ is a constant corresponding to the Schwarzschild mass.
The \enquote{usual} $(t, r, \theta, \varphi)$ coordinates are related to the coordinates $(u,r,z,\bar{z})$ by the change of variables
\begin{equation*}
	u = t - r - 2M \log(r - 2M) \qq{and} z = \cot\frac{\theta}{2} e^{i\varphi} \,.
\end{equation*}
In the following, we will keep working with the coordinates $(u,r,z,\bar{z})$. 

\subsection{Dynamics}

To describe the ASP of the Schwarzschild geometry, it is natural to introduce the perturbed $(P,L)$ which read 
\begin{equation}
	P(u, z, \bar{z}) = P_0 e^{\veps F_1} \,,\quad L(u, z, \bar{z}) = \veps L_1 \,,
\end{equation}
where $\epsilon$ is a bookkeeping parameter encoding the order of the expansion in perturbation\footnote{In principle, one could also consider a perturbation of the mass function $M$, but one can show that such a perturbation cannot be consistently introduced, such that one has to work with $M$ constant from the start.}.
Here, $F_1$ encodes the polar perturbations while $L_1$ encodes the axial perturbations.
%
It is further useful to define $N_1$ and $\Sigma_1$ by
\begin{equation}
	\Sigma = \veps \Sigma_1 + \mathcal{O}(\veps^2) \qq{and} N = \veps N_1 + \mathcal{O}(\veps^2) \,.
\end{equation}
Furthermore, we define the operator $\Delta_0$ by
\begin{equation}
	\Delta_0 = 2 P_0^2 \partial_z \partial_{\bar{z}} \,.
\end{equation}
One can note that $\Delta_0$ is such that $\Delta_0 \log P_0 = 1$ which corresponds to the Schwarzschild background value (i.e. to the purely spherically symmetric case). 

Consider the first dynamical equation (\ref{eq:evol-L}). Linearizing this equation gives
\begin{equation}
\label{EQ1}
	\partial_z N_1 = -3iM \partial_u L_1 \,,\qq{with} N_1 = \Sigma_1 + \frac12 \Delta_0\Sigma_1 \qq{and} \Sigma_1 = -P_0^2 \Im(\partial_z\bar{L}_1) \,.
\end{equation}
As expected, this equation for $L_1$ does not contain $F_1$ at first order in $\veps$ and the two axial and polar perturbations decouple. It is convenient to introduce a real function $f_1$ such that $L_1 = i \partial_z f_1$ which allows one to recast $(\Sigma, N)$ as
\begin{equation}
	\Sigma = \frac12 \Delta_0 f_1 \qq{and} N = \frac12 \Delta_0 f_1 + \frac14 \Delta_0\Delta_0 f_1 \,.
	\label{eq:perturb-f-M}
\end{equation}
With this new field $f_1$ to describe the axial perturbation, the equation (\ref{EQ1}) takes the form
\begin{equation}
\label{ee1}
	\boxed{\Delta_0 \Delta_0 f_1 + 2 \Delta_0 f_1 - 12 M \partial_u f_1 = 0} \,.
\end{equation}
It describes the dynamics of the linear axial ASP of the Schwarzschild background.

We can now turn to the second dynamical equation (\ref{eq:evol-P}). This equation being complex, it can be separated into two parts. The imaginary part yields
\begin{equation}
	\partial_u(\Delta_0 \Delta_0 f_1 + 2 \Delta_0 f_1 - 4 N_1) = 0 \,,
	\label{eq:pert-f}
\end{equation}
which is always verified given~\eqref{eq:perturb-f-M} while the real part gives
\begin{equation}
	\boxed{\Delta_0 \Delta_0 F_1 + 2 \Delta_0 F_1 + 12 M \partial_u F_1 = 0} \,.
	\label{eq:pert-F}
\end{equation}
It describes the dynamics of the linear polar ASP on the Schwarzschild black hole. Since both perturbations decouple at the linear order, if one focus solely on the linear polar sector, one can use the subset of Robinson-Trautman geometries to derive the same result. The dynamical equation (\ref{eq:pert-F}) is nothing else than the well known fourth order RT equation linearized around the Schwarzschild case $P= P_0$ \cite{Qi:1993ey}. 

Having derived the dynamical equations for the linear ASP of the Schwarzschild black hole, let us see how they can be analytically solved. These solutions are needed to construct the quadratic perturbations and in particular the source terms.

\subsection{Exact solutions}

%
A natural ansätze to solve these equations is to decompose the perturbations $(F_1, f_1)$ onto spherical harmonics defined by the eigenvalue equation
\begin{equation}
	\Delta_0 Y_{\ell}^m = - \ell(\ell+1) Y_{\ell}^m \,.
\end{equation}
More details about their definition are given in appendix~\ref{app:Ylm-stereo}. The ansätze reads
\begin{equation}
	F_1 (u, z, \bar{z})= \po{E} Y_{\po{\ell}}^{\po{m}} h_{\po{\ell}} (u) \qq{and} f_1(u, z, \bar{z}) = \ax{E} Y_{\ax{\ell}}^{\ax{m}} h_{\ax{\ell}}(u) \,,
	\label{eq:sol-f1-F1}
\end{equation}
where $\po{E}$ and $\ax{E}$ are constant amplitude factors. Notice that $\po{E}$ has no dimension while $\ax{E}$ is of dimension $M$.
Plugging this in equations (\ref{ee1}) and (\ref{eq:pert-F}) gives
\begin{align}
 &(\ax{\ell} - 1)\ax{\ell}(\ax{\ell}+1)(\ax{\ell}+2) h_{\ax{\ell}} - 12 M \dot{h}_{\ax{\ell}} = 0 \,,\\
 &(\po{\ell} - 1)\po{\ell}(\po{\ell}+1)(\po{\ell}+2) h_{\po{\ell}} + 12 M \dot{h}_{\po{\ell}} = 0 \,.
\end{align}
Defining $\kappa$ by
\begin{equation}
\label{k}
	\kappa(\ell) = \frac{(\ell - 1)\ell(\ell+1)(\ell+2)}{12M} \,,
\end{equation}
the general solutions for the ASP are given by
\begin{equation}
	F_1 = \po{E} Y_{\po{\ell}}^{\po{m}} e^{-\po{\kappa} u} \qq{and} f_1 = \ax{E} Y_{\ax{\ell}}^{\ax{m}} e^{+\ax{\kappa} u} \,,
	\label{eq:sol-f1-F1}
\end{equation}
where $\ax{\kappa} = \kappa(\ax{\ell})$ and $\po{\kappa} = \kappa(\po{\ell})$. One should note that, since perturbations $F_1$ and $f_1$ are decoupled, $\ax{\ell}$ and $\po{\ell}$ are different in general. If one inverts the time direction, the above solutions are mapped to 
\begin{equation}
	F_1 = \po{E} Y_{\po{\ell}}^{\po{m}} e^{+\po{\kappa} v} \qq{and} f_1 = \ax{E} Y_{\ax{\ell}}^{\ax{m}} e^{-\ax{\kappa} v} \,,
	\label{eq:sol-f1-F1}
\end{equation}
where $v$ is the advanced null coordinate. We have thus recovered the expressions for the linear ASP of the Schwarzschild black hole first derived in \cite{Couch:1973zc} and revisited more recently in \cite{Qi:1993ey}. These ASP reproduce the well-known solutions derived independently by Chandrasekhar in \cite{Chandrasekhar1984}.

\subsection{Zero modes}

\label{sec4}

With these exact solutions, we can now revisit the question of the zero modes. We refer the reader to \cite{Nakamura:2021mfv, Nakamura:2021ftr, Nakamura:2021jlc} for interesting discussion on their roles.
From (\ref{k}), we see that for $\ell=0$ and $\ell=1$, the ASP are static. Let us describe explicitly the two relevant cases.

First, consider the case where $\ax{\ell} =\po{\ell} =0$ which corresponds to a sum of axial and polar linear monopole perturbations. The metric is given by
\begin{align}
\rd s^2 = -  \left(  f - \frac{\epsilon \cE_p}{3} \right)  \rd u^2  - 2 \rd u \rd r - \left( 1 - \frac{\epsilon \cE_p}{3}\right) \frac{2r^2 \rd z \rd \bar{z}}{(1+ z\bar{z})^2}
\end{align}
where we have used the shorthand notation
\be
f = 1 - \frac{2M}{r} \,, \qquad \cE_{a,p} = \sqrt{\frac{3}{\pi}} E_{a,p} \,.
\ee
Notice that the $\ell=0$ axial zero mode does not appear in the solution, i.e. the metric does not depend on $E_a$. 
Changing the coordinates to
\begin{align}
 u & = t - r - 2M \log(r - 2M)\,, \\
 z & = \cot\frac{\theta}{2} e^{i\varphi}  \,,
\end{align}
leads to the metric 
\begin{align}
\rd s^2 = -f \left( 1 -  \frac{K}{f} \right) \rd t^2  - \frac{2 K }{f}  \rd t \rd r - f^{-1} \left( 1 - \frac{K}{f}\right) \rd r^2 + r^2 ( 1 + K) \rd \Omega^2 
\end{align}
where the constants $K$ are given by
\be
K 
= \frac{\epsilon \cE_p}{\sqrt{3}}   \,.
\ee
To remove the off diagonal terms and bring this metric to the Schwarzschild form, one can perform the following change of time coordinate given by
\begin{align}
t & = \left( 1 - \frac{K}{2}\right) \tilde{t} + \frac{\epsilon \cE_p}{\sqrt{3}} \left( \tilde{r} + 4M \log{(\tilde{r}-2M)}- \frac{4M^2}{\tilde{r}-2M} \right) \,,   \\
r & =\left( 1 + \frac{K}{2}\right) \tilde{r} \,,
\end{align}
which brings the metric to the standard Schwarzschild form
\begin{align}
\rd s^2 = - \left( 1 - \frac{2M(1 + \epsilon M_1)}{r} \right) \rd t^2 + \left( 1 - \frac{2M(1 + \epsilon M_1)}{r} \right)^{-1} \rd r^2 + r^2 \rd \Omega^2 \,,
\end{align}
where the shift in mass is given by
\be
M_1 = - \frac{3 K}{\sqrt{3}} \,.
\ee
As expected, the $\ell=0$ polar zero mode simply shifts the Schwarzschild mass.

Next, let us consider the zero mode corresponding to  $\ax{\ell}= \po{\ell} = 1$. The resulting geometry is stationary and given by
\be
\rd s^2 = - f \rd t^2 - 2 \epsilon \cE_a   \sin^2{\theta} \left( \frac{M}{r}  \rd t - \frac{\rd r}{2f}  \right)   \rd \varphi  +  \frac{\rd r^2}{f} + r^2 \left( 1- \epsilon \cE_p \cos{\theta} \right) \rd \Omega^2 \,.
\ee
Despite the non standard off diagonal terms and the presence of the three constants $(M, \cE_a, \cE_p)$, this line element describes the linearized Kerr black hole solution. To see that, one can apply the following change of the angular coordinates:
\begin{align}
\label{CCL}
\theta = \tilde{\theta} + \frac{1}{2} \epsilon  \cE_p \sin{\tilde{\theta}} \,, \qquad \qquad  \varphi = \tilde{\varphi} - \frac{\epsilon \cE_p}{4M} \log{f}  \,.
\end{align}
Within these new coordinates, the metric takes the more familiar form
\be
\label{SLOW}
\rd s^2 = - f \rd t^2 -  \frac{2 \epsilon \cE_a M }{r}  \sin^2{\tilde{\theta}} \rd t   \rd \tilde{\varphi}  +  \frac{\rd r^2}{f} + r^2  \rd \tilde{\Omega}^2 \,,
\ee
which one recognizes as being the well-known Kerr black hole geometry linearized w.r.t its angular momentum. We see that at the linear level, the dipolar axial perturbation introduces a (small) angular momentum controlled by its amplitude $\cE_a$. However, the dipolar polar perturbation can always be absorbed by the change of coordinates (\ref{CCL}) and the additional constant $\cE_p$ completely disappears, showing that it is a pure gauge, in agreement with the standard result \cite{Martel:2005ir}. 



\section{Quadratic algebraically special perturbations}

\label{sec5}

In this section, we generalize the results of Couch and Newman and we analyze the dynamics and present the analytic expression for the profiles of quadratic ASP of the Schwarzschild black hole.

\subsection{Dynamics}

Following the same procedure as for the linear perturbations, we now expand the two functions $(P,L)$ up to second order in perturbations
\begin{equation}
	P(u, z ,\bar{z}) = P_0 e^{\varepsilon F_1 + \varepsilon^2 F_2} \qq{and} L(u, z, \bar{z}) = \varepsilon L_1 + \varepsilon^2 L_2 \,.
\end{equation}
Furthermore, we introduce the function $f_2$ such that
\begin{equation}
	L_2 = i\partial_z f_2 + \ax{\kappa} \partial_z(f_1^2) \,.
\end{equation}
where we recall that $L_1$ is related to $f_1$ via $L_1 = i\partial_z f_1$. Now, using the expressions for the linear perturbations $(F_1, f_1)$, we expand up to second order the dynamical equations (\ref{eq:evol-L}) and (\ref{eq:evol-P}) which encdes the dynamics of the second order perturbations $(F_2, f_2)$.

After a lengthy computation, the dynamics for the second order axial perturbation $f_2$ splits into 
\begin{equation}
	\boxed{\Delta_0 \Delta_0 f_2 + 2 \Delta_0 f_2 - 12 M \partial_u f_2 = \mathcal{A} }\,,
	\label{eq:eq-f2}
\end{equation}
where $\A$ stands for the source term generated by the polar-axial mode interaction. Its expression reads
\begin{equation}
\label{sourceAP}
	\mathcal{A} = -2 F_1 \Delta_0\Delta_0 f_1 - 2 \Delta_0 F_1 \Delta_0 f_1 - 2 \Delta_0(F_1\Delta_0 f_1) - 8 F_1 \Delta_0 f_1 \,.
\end{equation}
After expanding up to second order the equation (\ref{eq:evol-P}), and having split its imaginary and real parts, the dynamical equation for the second order polar perturbation is given by
\begin{equation}
	\boxed{\Delta_0 \Delta_0 F_2 + 2 \Delta_0 F_2 + 12 M \partial_u F_2 = \mathcal{P} + \mathcal{Q} + \mathcal{R} }\,,
	\label{eq:eq-F2}
\end{equation}
where the source term is split into three different contributions given explicitly by
\begin{align}
\label{sourcePP}
	\mathcal{P} &= -2 \Delta_0(F_1\Delta_0 F_1) - 2 \Delta_0(F_1^2) - 2 F_1 \Delta_0\Delta_0 F_1 - 4 F_1 \Delta_0 F_1 \,, \\
	\label{sourceAA}
	\mathcal{Q} &= \ax{\kappa}^2 \Big[\Delta_0 (f_1^2) - f_1 \Delta_0\Delta_0f_1 + 4 (\Delta_0 f_1)^2 + 6 f_1 \Delta_0 f_1 + \frac32 \Delta_0\Delta_0(f_1^2) + 2 \Delta_0(f_1 \Delta_0 f_1)\Big] \,,\\
		\label{sourceAPP}
	\mathcal{R} &= 8 \po{\kappa} \Im[\partial_{\bar{z}}(P_0^2 \partial_{\bar{z}} F_1)\partial_{z}(P_0^2 \partial_{z} f_1)] - 8 (\ax{\kappa} - \po{\kappa}) \Im[P_0^2 \partial_{z} f_1 (\partial_{\bar{z}} \Delta_0 F_1 + 2 \partial_{\bar{z}} F_1)]\,.
\end{align}
It is direct to see that $\mathcal{P}$ encodes the polar-polar interaction, $\mathcal{Q}$ encodes the axial-axial interaction and finally $\mathcal{R}$ encodes the polar-axial interaction generating the source term. Let us further point that the differential operator acting on the second order perturbations $(F_2, f_2)$ is exactly the same as the one entering in the wave equation verified by the linear perturbations $(F_1, f_1)$. We now turn to the resolution of these equations.

\subsection{Exact solutions}

Despite the rather complicated form of these equations, let us show how one can construct exact analytical solutions for the quadratic ASP. This is the first main result of this work.

\subsubsection{Axial perturbations}

Consider first the quadratic axial ASP encoded in $f_2$. In order to solve Eqs.~\eqref{eq:eq-f2}, one can follow a procedure close to the one used to solve the linear perturbation equations. Given that $f_2$ is sourced only by axial-polar interactions, we can take the following ansatze:
\begin{equation}
	f_2 = \po{E}\ax{E} e^{(\ax{\kappa} - \po{\kappa})u} \sum_{\ell,m} a^\ell_m Y_\ell^m \,,
	\label{eq:sol-f2}
\end{equation}
where the $a^\ell_m$ are constant coefficients. These can be obtained from Eqs.~\eqref{eq:eq-f2} by making use of the decomposition of any product $Y_{\ell_1} Y_{\ell_2}$ onto the basis of $Y_\ell$. This decomposition uses the Clebsch-Gordan coefficients $k_{\ell_1 \ell_2 m}^{m_1 m_2 \ell}$, whose definition is given in App.~\ref{sec:decompo-coeffs} for convenience. Using this ansatze, the source term decomposes as
\begin{align}
	\mathcal{A} &= - \po{E}\ax{E} e^{(\ax{\kappa} - \po{\kappa})u} \Big[ 2 Y_{\po{\ell}}^{\po{m}} \Delta_0\Delta_0 Y_{\ax{\ell}}^{\ax{m}} + 2 \Delta_0 Y_{\po{\ell}}^{\po{m}} \Delta_0 Y_{\ax{\ell}}^{\ax{m}} + 2 \Delta_0(Y_{\po{\ell}}^{\po{m}}\Delta_0 Y_{\ax{\ell}}^{\ax{m}}) + 8 Y_{\po{\ell}}^{\po{m}} \Delta_0 Y_{\ax{\ell}}^{\ax{m}} \Big] \,,\nn\\
	&= \po{E}\ax{E} e^{(\ax{\kappa} - \po{\kappa})u} \sum_\ell \mathcal{A}_m^\ell Y_\ell^m \,,
\end{align}
where the coefficients $\mathcal{A}_m^\ell$ are given explicitly by
\begin{equation}
	\mathcal{A}_m^\ell = -8 (\ax{\lambda} +1) (\po{\lambda} + \ax{\lambda} + \lambda + 1) k^{\ax{m}\po{m}\ell}_{\ax{\ell}\po{\ell}m} \,,
	\label{eq:expr-Alm}
\end{equation}
In these expressions, $\lambda$ is defined from $\ell$ (and similarly, $\ax{\lambda}$ and $\po{\lambda}$ from $\ax{\ell}$ and $\po{\ell}$ respectively) using Eq.~\eqref{eq:def-lambda}. Using these expressions and plugging them in the dynamical equation \eqref{eq:eq-f2}, one obtains the following constraint 
\begin{equation}
	\sum_{\ell,m} \Big[(\ell-1)\ell(\ell+1)(\ell+2) a^\ell_m -12M (\ax{\kappa} - \po{\kappa}) a^\ell_m - \mathcal{A}^\ell_m \Big] Y_\ell^m = 0 \,.
\end{equation}
which holds for all $\ell$.
Since the $Y_\ell^m$ form a basis of the space of real functions on the 2-sphere, every coefficient must be put to zero, leading to 
\begin{equation}
\label{sol-f2}
	12M (\kappa - \ax{\kappa} + \po{\kappa}) a^\ell_m - \mathcal{A}^\ell_m = 0 \,.
\end{equation}
If $\ax{\kappa} \neq \po{\kappa}$, this equation can be solved by imposing
\begin{equation}
	a^\ell_m = \frac{1}{12M} \frac{\mathcal{A}^\ell_m}{\kappa - \ax{\kappa} + \po{\kappa} } \,.
\end{equation}
If one has $\kappa = \ax{\kappa} - \po{\kappa}$ then the only solution is $a^\ell_m = 0$, except when $\ell = 0$ or $\ell = 1$. The question oof the zero modes will be discussed at the end of the section.

\subsubsection{Polar perturbations}

Let us now consider the quadratic polar perturbations encoded in $F_2$. In this case, the source term is much more involved. Nevertheless, a good ansatze to this equations 
is given by
\begin{equation}
	F_2 = \po{E}^2 e^{-2\po{\kappa}u} \sum_{\ell,m} p^\ell_m Y_\ell^m + \ax{E}^2e^{2\ax{\kappa}u} \sum_{\ell,m} q^\ell_m Y_\ell^m + \po{E}\ax{E} e^{(\ax{\kappa} - \po{\kappa})u} \sum_{\ell,m} r^\ell_m Y_\ell^m \,,
	\label{eq:sol-F2}
\end{equation}
with $p^\ell_m$, $q^\ell_m$ and $r^\ell_m$ constant coefficients. Decomposing the source terms $\mathcal{P}$ and $\mathcal{Q}$, one obtains the following expression for $\mathcal{P}$
\begin{align}
	\mathcal{P} &= - 2 \po{E}^2 e^{-2\po{\kappa}u} \Big\{\Big[ -\po{\ell}(\po{\ell}+1) + 1 \Big]\Delta_0(Y_{\po{\ell}}^2) + \Big[\po{\ell}^2(\po{\ell}+1)^2 - 2 \po{\ell}(\po{\ell}+1)\Big]Y_{\po{\ell}}^2 \Big\} \,, \nn \\
	&= \po{E}^2 e^{-2\po{\kappa}u} \sum_{\ell,m} \mathcal{P}^\ell_m Y_\ell^m \,.
\end{align}
while $\mathcal{Q}$ is given by
\begin{align}
	\mathcal{Q} &= \ax{\kappa}^2 \ax{E}^2 e^{2\ax{\kappa}u} \Big\{ \frac32\Delta_0\Delta_0(Y_{\ax{\ell}}^2) + \Big[-2 \ax{\ell}(\ax{\ell}+1) + 1\Big]\Delta_0(Y_{\ax{\ell}}^2) + \Big[3\ax{\ell}^2(\ax{\ell}+1)^2 - 6 \ax{\ell}(\ax{\ell}+1) \Big]Y_{\ax{\ell}}^2 \Big\}\,,\nn \\
	&= \ax{E}^2 e^{2\ax{\kappa}u} \sum_{\ell,m} \mathcal{Q}^\ell_m Y_\ell^m \,.
\end{align}
In both expressions, the coefficients $\mathcal{P}^\ell_m$ and $\mathcal{Q}^\ell_m$ are defined by
\begin{align}
	\mathcal{P}^\ell_m &= -4 \Big[1 + \lambda + 4 \po{\lambda} + 2\lambda\po{\lambda} + 2 \po{\lambda}^2\Big] k^{\po{m}\po{m}\ell}_{\po{\ell}\po{\ell}m} \,,\label{eq:expr-Plm}\\
	\mathcal{Q}^\ell_m &= 2\ax{\kappa}^2 \Big[6 + 3\lambda(\lambda+3) + 10\ax{\lambda} + 4\lambda\ax{\lambda} + 6 \ax{\lambda}^2\Big] k^{\ax{m}\ax{m}\ell}_{\ax{\ell}\ax{\ell}m} \,.\label{eq:expr-Qlm}
\end{align}
In order to decompose the source term $\mathcal{R}$, it is convenient to introduce the coefficients $h$ and $l$ defined in App.~\ref{sec:decompo-coeffs}. Using these definitions, one can write
\begin{equation}
	\mathcal{R} = \po{E} \ax{E} e^{(\ax{\kappa}-\po{\kappa})u} \sum_{\ell,m} \mathcal{R}^\ell_m Y_\ell^m \,,
	\label{eq:decompo-R}
\end{equation}
where the coefficients $\mathcal{R}^\ell_m$ reads
\begin{equation}
	\mathcal{R}^\ell_m = 8\po{\kappa}\Im\big(l_{\ax{\ell}\po{\ell}m}^{\ax{m}\po{m}\ell}\big) + 16 \po{\lambda} (\ax{\kappa} - \po{\kappa})\Im\big( h_{\ax{\ell} \po{\ell} m}^{\ax{m} \po{m}  \ell}\big)  \,.
\end{equation}
In the case $\ax{m} = \po{m} = 0$, one has $\mathcal{R} = 0$, because $h_{\ax{\ell}\po{\ell}m}^{00\ell}$ and $l_{\ax{\ell}\po{\ell}m}^{00\ell}$ are reals (see App.~\ref{app:Ylm-stereo}). This means that for $\ax{m} = \po{m} = 0$, quadratic polar perturbations cannot be sourced by axial-polar interaction; this is expected for parity conservation reasons. 

Using the above decomposition of the source terms, the dynamical equation \eqref{eq:eq-F2} for the quadratic polar perturbations splits into three conditions given by
\begin{align}
	&\sum_\ell\Big[(\ell-1)\ell(\ell+1)(\ell+2) p^\ell_m - 24 M \po{\kappa} p^\ell_m - \mathcal{P}^\ell_m\Big] Y_\ell = 0 \,,\\
	&\sum_\ell\Big[(\ell-1)\ell(\ell+1)(\ell+2) q^\ell_m + 24 M \ax{\kappa} q^\ell_m - \mathcal{Q}^\ell_m\Big] Y_\ell = 0 \,,\\
	&\sum_\ell\Big[(\ell-1)\ell(\ell+1)(\ell+2) r^\ell_m + 12 M (\ax{\kappa} - \po{\kappa}) r^\ell_m - \mathcal{R}^\ell_m\Big] Y_\ell = 0 \,.
\end{align}
These equations are solved by taking
\begin{equation}
\label{sol-F2}
	p^\ell_m = \frac{1}{12M} \frac{\mathcal{P}^\ell_m}{\kappa - 2 \po{\kappa}}\,, \quad q^\ell_m = \frac{1}{12M} \frac{\mathcal{Q}^\ell_m}{\kappa + 2 \ax{\kappa}} \,, \quad r^\ell_m = \frac{1}{12M} \frac{\mathcal{R}^\ell_m}{\kappa + \ax{\kappa} - \po{\kappa}} \,.
\end{equation}
These expressions provide an analytic exact solutions for the quadratic ASP of the Schwarzschild black hole.

In order to demonstrate the usefulness of the exact analytical solutions presented here, we have computed the profiles of the quadratic perturbations $(F_2, f_2)$ sourced by linear perturbations $(F_1, f_1)$ with quantum numbers $\ax{\ell} = 2$, $\ax{m} = 1$ and $\po{\ell} = 3$, $\po{m} = 2$. The detailed derivation is presented in appendix~\ref{examp}. This solution describes the quadratic perturbation sourced by the interaction of a linear axial quadrupole with a linear polar hexapole which have a decay time given respectively by
\begin{equation}
\label{ex-f2}
	\ax{\kappa} = \frac{2}{M} \,,\quad \po{\kappa} = \frac{10}{M}
\end{equation}
The analytic profile of the quadratic axial ASP sourced by the polar-axial interaction is given by
\begin{equation}
\label{ex-F2}
	f_2 = \po{E}\ax{E} \Big(-\sqrt{\frac{3}{7\pi}} Y_1^1 - \frac{13}{18\sqrt{6\pi}} Y_3^1 - \frac{13}{54}\sqrt{\frac{5}{2\pi}} Y_3^3 + \frac{2}{39}\sqrt{\frac{55}{21\pi}} Y_5^1 - \frac{2}{117}\sqrt{\frac{110}{\pi}} Y_5^3\Big) e^{-8u/M} \,.
\end{equation}
while the quadratic polar ASP has additional contributions coming from the polar-polar, axial-axial and axial-polar interactions. It reads
\begin{align}
	F_2 & = \po{E}^2 \Big(\frac{1}{2\sqrt{\pi}} Y_0^0 + \frac{119}{66\sqrt{\pi}} Y_4^0 - \frac{17}{66}\sqrt{\frac{35}{\pi}} Y_4^4 - \frac{97}{264\sqrt{13\pi}} Y_6^0 - \frac{97}{264}\sqrt{\frac{7}{13\pi}} Y_6^4\Big) e^{-20u/M} \\ 
	& + \frac{\ax{E}^2}{M^2} \Big(\frac{3}{\sqrt{\pi}} Y_0^0 + \frac{16}{21}\sqrt{\frac{5}{\pi}} Y_2^0 + \frac{16}{7} \sqrt{\frac{5}{3\pi}} Y_2^2 - \frac{892}{357\sqrt{\pi}} Y_4^0 + \frac{446}{357}\sqrt{\frac{5}{\pi}} Y_4^2\Big) e^{4u/M} \\ 
	& + \frac{\po{E}\ax{E}}{M} \Big(\frac{25}{33} \sqrt{\frac{5}{2\pi}} Y_4^{-3} - \frac{25}{33} \sqrt{\frac{35}{2\pi}} Y_4^{-1}\Big) e^{-8u/M} \,.
\end{align}
Notice that the contribution coming from the axial-polar interaction is non-vanishing because we have chosen linear ASP with non-vanishing magnetic numbers. 

\subsection{Quadratic zero modes}

Now, we can use our exact solutions for both polar and axial quadratic ASP to analyze the quadratic zero modes. From (\ref{eq:sol-f2}) and (\ref{eq:sol-F2}), we see that in order to have time-independent quadratic perturbations, one has to fix
\be
\ax{\kappa} = \po{\kappa} =0
\ee
which further select the time-independent linear perturbations. 
They are of three types:
\begin{itemize}
\item $\ax{\ell} = \po{\ell} =1$
\item $\ax{\ell} = 0$ and $ \po{\ell} =1$
\item $\ax{\ell} = 1$ and  $\po{\ell} = 0$
\end{itemize}
For each of them, three subcases have to be distinguish depending on the values of the magnetic numbers $\ax{m}$ and $\po{m}$.

\subsubsection{Monopole-monopole zero mode }

Consider the case where $\ax{\ell} =\po{\ell} =0$. The associated metric is given by
\begin{align}
\rd s^2 = -f \left( 1 -  \frac{K_{+}}{f} \right) \rd t^2  + \frac{2 K_{+} }{f}  \rd t \rd r - f^{-1} \left( 1 - \frac{K_{+}}{f}\right) \rd r^2 + r^2 ( 1 - K_{-}) \rd \Omega^2 
\end{align}
where the constants $K_{\pm}$ are given by
\be
K_{\pm} = \frac{\epsilon \cE_p}{\sqrt{3}}    \left(  \frac{\epsilon \cE_p}{2 \sqrt{3}} \pm 1 \right)
\ee
To compare this to the Schwarzschild metric, we need to remove the off diagonal terms. This is achieved by performing the change
\begin{align}
\label{diffSct}
t & = \left( 1 + \frac{K_{-}}{2}\right) \tilde{t} + \frac{\epsilon \cE_p}{\sqrt{3}} \left( \tilde{r} + 4M \log{(\tilde{r}-2M)}- \frac{4M^2}{\tilde{r}-2M} \right)  + \frac{4 \epsilon^2 \cE^2_p M}{\sqrt{3}} \left( \frac{2M (2 \tilde{r} - 3M)}{(\tilde{r}-2M)^2}- \log{(\tilde{r}-2M)} \right)  \\
\label{diffScr}
r & =\left( 1 + \frac{K_{+}}{2}\right) \tilde{r} 
\end{align}
which bring the metric to the standard Schwarzschild form
\begin{align}
\label{SchQ}
\rd s^2 = - \left( 1 - \frac{2M(1 + \epsilon M_1 + \epsilon^2 M_2)}{r} \right) \rd t^2 + \left( 1 - \frac{2M(1 + \epsilon M_1 + \epsilon^2 M_2)}{r} \right)^{-1} \rd r^2 + r^2 \rd \Omega^2
\end{align}
The new constants $(M_1, M_2)$ are given in term of the polar perturbation by
\be
\label{SchQM}
M_1 = - \frac{3 }{2\sqrt{3}} \cE_p \qq{and} M_2 = \frac{3 }{8} \cE^2_p \,.
\ee
As expected, it corresponds to a quadratic shift of the Schwarzschild mass.

\subsubsection{Dipole-dipole zero mode}

Let us focus on the simplest one corresponding to
$\ax{\ell} = \po{\ell} =1$ and $\ax{m} = \po{m} =0$. This geometry does not depend on time nor on the angular coordinate $\varphi$ since the magnetic numbers have been chosen to vanish. It is therefore axi-symmetric and the associated metric is given by
\begin{align}
\label{metsee}
\rd s^2 & = \left( 1 - \frac{ K_1(r, \theta)}{ r^3 f}  \right) \left(  - f  \rd t^2 +  \frac{ \rd r^2}{f} \right)  +    \frac{ 2 K_1(r, \theta)}{ r^3 f}   \rd t \rd r + 2 K_4 (r,\theta)\left( \frac{M}{r}  \rd t - \frac{\rd r}{2f}  \right)\rd \varphi  \\
& \;\;\; +  r^2 \left( K_2(r,\theta) \rd \theta^2 + K_3 (r, \theta) \sin^2{\theta} \rd \varphi^2\right) 
\end{align}
where the four functions are given by
 \begin{align}
K_1(r, \theta) & = \frac{1}{4} \left( \epsilon^2  \cE^2_a M (1+ \cos{2\theta}) - \frac{2}{3}\epsilon^2  \cE^2_p r^3 \right) \\
K_2(r, \theta) & = 1- \epsilon \cE_p \cos{\theta} + \frac{\epsilon^2}{8} \left( \cE^2_a + \frac{7}{3} \cE^2_p r^2 + (\cE^2_a + 3 \cE^2_p r^2) \cos{2\theta}\right) \\
K_3(r,\theta) & = 1 -  \epsilon \cE_p \cos{\theta}  + \frac{\epsilon^2}{8} \left( \frac{7}{3} \cE^2_p r^3 + 2 \cE^2_a (M +r) + (\cE^2_p r^3 - 2 \cE^2_a M) \cos{2\theta} \right) \\
K_4(r,\theta) & = - \left( 1 - \epsilon  \cE_p   \cos{\theta} \right)  \epsilon  \cE_a M  \sin^2{\theta}
 \end{align}
By virtue of the no-hair theorem, this metric is expected to be related to the slowly rotating Kerr metric up to quadratic order in the spin. Let us write down the explicit diffemorphism which allows to transform the above metric to the more familiar form. To proceed, we introduce the following change of coordinates 
\begin{align}
\label{diffKerr1}
t & = \left(1 + \frac{\epsilon^2 \cE^2_p}{12}\right) \tilde{t} + \frac{\epsilon^2}{8} \left[ \frac{2\tilde{r}}{f(\tilde{r})} \left(\frac{8}{3}\cE^2_p M^2 -  \cE^2_a \right) - \frac{ \cE^2_a}{M} \log{f(\tilde{r})} - \frac{16}{3} \cE^2_p M \log{(\tilde{r}-2M)}  - \frac{\cE^2_p \tilde{r}}{12}  \right] \,,\\
r & = \left(1 - \frac{\epsilon^2 \cE^2_p}{12}\right) \tilde{r}  \,,\\
\theta & = \tilde{\theta} + \frac{1}{2} \epsilon  \cE_p \sin{\tilde{\theta}} + \frac{ \epsilon^2 }{16} \cE^2_p \sin{2\tilde{\theta}} \,,\\
\label{diffKerr2}
  \varphi & = \tilde{\varphi} - \frac{\epsilon \cE_a}{4M} \log{f(\tilde{r})}  \,. 
\end{align}
Within these new coordinates, the off diagonal $g_{tr}$ and $g_{r\varphi}$ in (\ref{metsee}) disappear and the metric takes the following form
\begin{align}
\rd s^2 & = - \left[ f(\tilde{r}) + \epsilon^2 \left( \frac{2M_2}{r} - \frac{2 a M \cos^2{\tilde{\theta}}}{\tilde{r}^3 } \right) \right] \rd \tilde{t}^2  +\left[  f^{-1}(\tilde{r}) + \frac{\epsilon^2}{\tilde{r} f(\tilde{r})}  \left( \frac{a^2 - 2\tilde{r} M_2}{f(\tilde{r})} + a^2  \cos^2{\tilde{\theta}} \right) \right] \rd \tilde{r}^2 \nn  \\
\label{KerrQ}
& - \frac{4 \epsilon a M \sin^2{\tilde{\theta}}}{\tilde{r}} \rd \tilde{t} \rd \tilde{\varphi} + \left( \tilde{r}^2 + \epsilon^2 a^2 \cos^2{\tilde{\theta}} \right) \rd \tilde{\theta}^2 + \left[ 1 + \frac{\epsilon^2 a^2 (\tilde{r} + 2 M \sin^2{\tilde{\theta}})}{\tilde{r}^3}\right]  \tilde{r}^2 \sin^2{\tilde{\theta}}\rd \tilde{\varphi}^2  \,,
\end{align}
where the constants $(a, \cM, M_2)$ are given in terms of the perturbations by
\be
\label{SpinQ}
 M_2 = \frac{\cE_p^2 M }{4} \qq{and}  a = \frac{\cE_a}{2}   \,.
\ee
and where $M$ is the original Schwarzschild mass.
As expected, it corresponds to the slowly rotating Kerr metric up to quadratic order in the spin $a$. We see that the Schwarzschild mass is shifted by the polar perturbation, while the spin is generated by the axial perturbation.
This geometry admits a new Killing horizon located at 
\be
r_{+} = 2M  +  \frac{\epsilon^2 M}{3} \left( \cE^2_p - \frac{\cE^2_a}{8M^2} \right)
\ee
Therefore, contrary to the linear regime where the horizon of the black hole perturbed by the dipolar zero mode is not shifted, in the non-linear regime, the Schwarzschild black hole geometry dressed with the quadratic zero modes admits a shifted horizon at $r_{+} \neq 2M$.  

\section{Discussion}

\label{sec6}

In this work, we have revisited the description of the algebraically special perturbations (ASPs) of the Schwarzschild black hole and extended the analysis of Couch and Newman to the non-linear regime. The dynamic of the quadratic ASPs can be derived by expanding up to second order the reduced Einstein equations (\ref{eq:evol-L} -- \ref{eq:evol-P}) of the most general vacuum twisting family of solutions of general relativity. Expanding these equations up to second order in the perturbations of the two fields $(P,L)$ which label this family, we have obtained the master non-linear wave equations (\ref{eq:eq-f2}) and (\ref{eq:eq-F2}) which describe respectively the dynamic of the quadratic axial perturbations $f_2$ and the dynamic of quadratic polar perturbations $F_2$. By construction, these solutions of these equations preserve the Petrov type II of the geometry. 

A nice feature of these equations is that the sources can be written explicitly. From these expressions, we see that the quadratic axial perturbation $f_2$ is sourced by the axial-polar interaction encoded in the source $\A$ given by (\ref{sourceAP}), while the quadratic polar perturbation $F_2$ is sourced by three types of interactions: polar-polar~\eqref{sourcePP}, axial-axial~\eqref{sourceAA} and axial-polar~\eqref{sourceAPP}. This last contribution is only present when the involved linear perturbations exhibit non-zero magnetic numbers. 

Then, we have shown that upon using the ansätze (\ref{eq:sol-f2}) and (\ref{eq:sol-F2}) together with a suitable decomposition of the sources in terms of spherical harmonics, one can construct analytic solutions to these inhomogeneous wave equations upon solving the condition (\ref{sol-f2}) for the axial perturbations and the relations (\ref{sol-F2}) for the polar perturbations. We have further presented an explicit example, given by (\ref{ex-f2}) and (\ref{ex-F2}),  of an analytic solution corresponding to a quadratic ASP sourced by a linear axial quadrupole and a linear polar hexapole. This example demonstrates that i) the damping time of the quadratic ASPs is related as expected to the damping times of the linear perturbations and ii) that the known divergence at the horizon of the linear ASPs is not cured by their non-linear interactions.

Finally, we have treated the case of the quadratic zero modes, focusing on the $\ell=0$ and $\ell=1$ quadratic ASPs. We have shown that as expected, the quadratic zero mode sourced by the $\ell=0$ axial and polar linear modes corresponds to a Schwarzschild black hole with quadratic corrections of its mass given by (\ref{SchQ}). These corrections are sourced solely by the polar mode through its amplitude $\cE_p$. Similarly, we have shown that the quadratic zero mode sourced by the $\ell=1$ axial and polar zero linear modes (with $\ax{m}=\po{m}=0$) deform the Schwarzschild background to the Kerr geometry up to quadratic corrections in the spin and in the mass corresponding to the metric (\ref{KerrQ}). As expected, the spin correction (\ref{SpinQ}) is generated by the axial sector and related to the amplitude $\cE_a$. While these results are expected, we provide here the explicit diffeomorphisms one has to implement to confirm these results. There are given by (\ref{diffSct} - \ref{diffScr}) and (\ref{diffKerr1} - \ref{diffKerr2}). To our knowledge, this has not been presented in that form elsewhere.

This study can serves as the basis to explore several open questions. First, it would be interesting to determine if, similarly to the linear case \cite{MaassenvandenBrink:2000iwh}, one can find well-behaved highly damped QQNMs of the Schwarzschild black hole which are spectrally close (i.e. numerically indistinguishable) from the quadratic ASPs described in this work. Second, one can use the expression for the quadratic ASP to search for possible hidden symmetry structures  in the non-linear regime of black hole perturbations. Finally, we stress that while pathological, the ASPs provide one of the few examples corresponding to explicit solutions in the time-domain. Finding other exact solutions within a closed form in the time-domain but describing instead well-behaved perturbations remains an interesting goal, as it would surely allows one to better understand the different phases of the signal during the different stages of the relaxation process.

\newpage
\appendix

\section{Spherical harmonics in stereographic coordinates}
\label{app:Ylm-stereo}

In this appendix, we present the relevant properties of the spherical harmonics which we use throughout this work.

\subsection{Definition}

Spherical harmonics are eigenfunctions of the Laplacian operator $\Delta_0$:
\begin{equation}
	\Delta_0 Y_\ell^m = -\ell(\ell+1) Y_\ell^m \,.
	\label{eq:def-Ylm-eigenvalues}
\end{equation}
They are indexed by the two integers $\ell$ and $m$, with $\ell \in \mathbb{N}$ and $m \in \llbracket -\ell, \ell \rrbracket$. The most common way to construct them is the following:
\begin{equation}
	\underline{Y}_\ell^m = (-1)^m \sqrt{ \frac{2\ell+1}{4\pi} \frac{(\ell-m)!}{(\ell+m)!}} P_\ell^m\Big(\frac{z\bar{z} - 1}{z\bar{z} + 1}\Big) \Big(\frac{z}{\bar{z}}\Big)^{m/2} \,.
	\label{eq:def-Ylm-complex}
\end{equation}
Here, $P_\ell^m$ represent an associated Legendre polynomial, and we have written the eigenfunctions $\underline{Y}$ to emphasize the fact that they take their values in $\mathbb{C}$.

As defined in~\eqref{eq:def-Ylm-complex}, spherical harmonics satisfy the following properties beyond the eigenvalue equation~\eqref{eq:def-Ylm-eigenvalues}:
\begin{align}
	&\Big(z \pdv{}{z} - \bar{z} \pdv{}{\bar{z}}\Big) \underline{Y}_\ell^m = m \underline{Y}_\ell^m \,,\\
	&\big(\underline{Y}_\ell^m\big)^* = (-1)^m \underline{Y}_\ell^{-m} \,,\\
	&(1 + z\bar{z}) \partial_z \underline{Y}_\ell^m = \frac{m}{z} \underline{Y}_\ell^m + \frac{\bar{z}}{z} \sqrt{(\ell-m)(\ell + m + 1)} \underline{Y}_\ell^{m+1} \,,\\
	&(1 + z\bar{z}) \partial_{\bar{z}} \underline{Y}_\ell^m = - m z \underline{Y}_\ell^m + \sqrt{(\ell-m)(\ell + m + 1)} \underline{Y}_\ell^{m+1} \,,\\
	&\int \frac{2\dd{z}\dd{\bar{z}}}{P_0^2} \underline{Y}_\ell^m \big(\underline{Y}_{\ell'}^{m'}\big)^* = \delta_{\ell\ell'} \delta_{mm'}  \,.
\end{align}

\subsection{Real spherical harmonics}

In this work, we focus on \emph{real} spherical harmonics. These are built from the complex harmonics $\underline{Y}_\ell^m$ by
\begin{equation}
	Y_\ell^m = \begin{cases*}
		\frac{1}{\sqrt{2}} \big[\underline{Y}_\ell^m + (-1)^m \underline{Y}_\ell^{-m}\big] = \sqrt{2} \Re(\underline{Y}_\ell^m) & if $m > 0$, \\
		\underline{Y}_\ell^0 & if $m = 0$, \\
		\frac{1}{\sqrt{2}i} \big[\underline{Y}_\ell^\abs{m} - (-1)^\abs{m} \underline{Y}_\ell^{-\abs{m}}\big] = \sqrt{2} \Im(\underline{Y}_\ell^\abs{m}) & if $m < 0$. \\
	\end{cases*}
\end{equation}
These harmonics still satisfy the orthonormality relation:
\begin{equation}
	\int \frac{2\dd{z}\dd{\bar{z}}}{P_0^2} Y_\ell^m Y_{\ell'}^{m'} = \delta_{\ell\ell'} \delta_{mm'} \,.
\end{equation}
In table~\ref{tab:Ylm}, we give the expression of the first real spherical harmonics in stereographic coordinates.
{\renewcommand{\arraystretch}{2.5}
\begin{table}[!htb]
	\begin{tabular}{|>{\centering\arraybackslash}p{1cm}|c||M{5cm}|}
		\hline
		$\bm{\ell}$ & $\bm{m}$& \bm{Y_\ell^m} \\\hline
		0 & 0 & \frac{1}{2\sqrt{\pi}} \\\hline
		\multirow{3}*{1} & -1 & \sqrt{\frac{3}{\pi}} \frac{\Im(z)}{1+z\bar{z}} \\
		& 0 & \sqrt{\frac{3}{\pi}} \frac{-1 + z\bar{z}}{2(1+z\bar{z})} \\
		& 1 & \sqrt{\frac{3}{\pi}} \frac{\Re(z)}{1+z\bar{z}} \\\hline
		\multirow{5}*{2} & -2 &  \sqrt{\frac{15}{\pi}} \frac{\Im(z^2)}{(1+z\bar{z})^2}\\
		 & -1 &  \sqrt{\frac{15}{\pi}} \Im(z)\frac{-1 + z\bar{z}}{(1+z\bar{z})^2}\\
		 & 0 &  \frac12 \sqrt{\frac{5}{\pi}} \frac{1 - 4z\bar{z} + z^2\bar{z}^2}{(1 + z\bar{z})^2}\\
		 & 1 &  \sqrt{\frac{15}{\pi}} \Re(z)\frac{-1 + z\bar{z}}{(1+z\bar{z})^2}\\
		 & 2 &  \sqrt{\frac{15}{\pi}} \frac{\Re(z^2)}{(1+z\bar{z})^2}\\
		\hline
	\end{tabular}
	\caption{Expressions of the first $Y_\ell^m$.}
	\label{tab:Ylm}
\end{table}
}

\subsection{Decomposition coefficients}
\label{sec:decompo-coeffs}

Several times throughout this paper, it is necessary to decompose a product of $Y_\ell^m$ functions or their derivatives onto a basis of functions. Here, we define all the coefficients we use. They can be computed using the scalar product $\braket{\cdot}{\cdot}$ defined by
\begin{equation}
	\braket{f}{g} = \int \frac{2\dd{z}\dd{\bar{z}}}{P_0^2} f g \,.
\end{equation}
This is possible because the real spherical harmonics form an orthonormal basis of functions on the 2-sphere.
\begin{itemize}
	\item The most important coefficients are the ones describing the decomposition of a product of two spherical harmonics onto the basis of spherical harmonics:
	\begin{equation}
		Y_{\ell_1}^{m_1} Y_{\ell_2}^{m_2} = \sum_{\ell,m} k^{m_1 m_2 \ell}_{\ell_1 \ell_2 m} Y_\ell^m \qq{with} k^{m_1 m_2 \ell}_{\ell_1 \ell_2 m} = \braket{Y_\ell^m}{Y_{\ell_1}^{m_1} Y_{\ell_2}^{m_2}} \,.
	\end{equation}
	These coefficients are the well-known Clebsch-Gordan coefficients.
	\item In order to compute the coefficients $\mathcal{R}_m^\ell$ in~\eqref{eq:decompo-R}, the metric expansions in App.~\ref{app:exp-metric} or the metric reconstruction in Sec.~\ref{sec:metric-reconstruction}, one needs to use the coefficients $h$ defined by
	\begin{equation}
		P_0^2 \partial_z Y_{\ell_1}^{m_1} \partial_{\bar{z}} Y_{\ell_2}^{m_2} = \sum_{\ell,m} h_{\ell_1 \ell_2 m}^{m_1 m_2 \ell} Y_\ell^m \qq{or} h_{\ell_1 \ell_2 m}^{m_1 m_2 \ell} = \braket{Y_\ell^m}{P_0^2 \partial_z Y_{\ell_1}^{m_1} \partial_{\bar{z}} Y_{\ell_2}^{m_2}} \,.
	\end{equation}
	These coefficients are complex when $(\ell_1, m_1) \neq (\ell_2, m_2)$: in general, one has
	\begin{equation}
		\big(h_{\ell_1 \ell_2 m}^{m_1 m_2 \ell}\big)^* = h_{\ell_2 \ell_1 m}^{m_2 m_1 \ell} \,.
	\end{equation}
	One should note that in the case $m_1 = m_2 = 0$, these coefficients are real.
	\item The computation of the coefficients $\mathcal{R}^\ell_m$ also requires the coefficients $l$ such that
	\begin{equation}
		\partial_{z}(P_0^2 \partial_{z} Y_{\ell_1}^{m_1}) \partial_{\bar{z}}(P_0^2 \partial_{\bar{z}} Y_{\ell_2}^{m_2}) = \sum_{\ell,m} l_{\ell_1 \ell_2 m}^{m_1 m_2 \ell} Y_\ell^m \,.
	\end{equation}
	These coefficients can be computed using
	\begin{equation}
		l_{\ell_1 \ell_2 m}^{m_1 m_2 \ell} = \braket{Y_\ell^m}{\partial_{\bar{z}}(P_0^2 \partial_{\bar{z}} Y_{\ell_1}^{m_1}) \partial_z(P_0^2 \partial_z Y_{\ell_2}^{m_2})} \,.
	\end{equation}
	One should note that in the case $m_1 = m_2 = 0$, these coefficients are real.

	\item Finally, the metric reconstruction also requires defining the coefficients $j$ such that
	\begin{equation}
		\big(\partial_z Y_{\ell_1}^{m_1}\big)^2 = \sum_{\ell, m} j_{\ell_1 m}^{m_1 \ell} (\partial_z^2 + 2 \partial_z(\log P_0) \partial_z) Y_\ell^m \,,
	\end{equation}
	and the coefficients $g$ defined by
	\begin{equation}
		Y_{\ell_1}^{m_1} \partial_z Y_{\ell_2}^{m_2} = \sum_{\ell, m} g_{\ell_1 \ell_2 m}^{m_1 m_2 \ell} \partial_z Y_\ell^m \,.
	\end{equation}
\end{itemize}
With these definitions, one can decompose the various appearance of product of spherical harmonics and their derivatives in a straigtforward way.

\section{Reconstruction of the perturbed geometry}
\label{app:exp-RW}

In this appendix, we present the reconstruction of the full metric up to second order. We also compute of the metric perturbation functions as defined by Regge and Wheeler~\cite{Regge:1957td}.

\subsection{Perturbative expansion of the twisting vacuum radiative metric}
\label{app:exp-metric}

Let us first compute the $\mathcal{O}(\varepsilon)$ and $\mathcal{O}(\varepsilon^2)$ contributions to the metric perturbations in the $(u,r, z, \bar{z})$ coordinates. The expansions of the functions appearing in~\eqref{eq:def-funcs-metric} are written as
\begin{align}
	L&= \varepsilon L_1 + \varepsilon^2 L_2 + \mathcal{O}(\varepsilon^3) \,, & P &= P_0 e^{\varepsilon F_1 + \varepsilon^2 F_2} + \mathcal{O}(\varepsilon^3) \,,\\
	\Sigma &= \varepsilon \Sigma_1 + \varepsilon^2 \Sigma_2 + \mathcal{O}(\varepsilon^3) \,, &K &= 1 + \varepsilon K_1 + \varepsilon^2 K_2 + \mathcal{O}(\varepsilon^3) \,, \\
	W &= \varepsilon W_1 + \varepsilon^2 W_2 + \mathcal{O}(\varepsilon^3) \,, & N &= \varepsilon N_1 + \varepsilon^2 N_2 + \mathcal{O}(\varepsilon^3) \,, \\
	2H &= A(r) + 2 \varepsilon H_1 + 2 \varepsilon^2 H_2 + \mathcal{O}(\varepsilon^3)\,,
\end{align}
where the linear contributions are given by
\begin{align}
	L_1 &= i \partial_z f_1 \,, & P_1 &= P_0 F_1 \,, &\Sigma_1 &= \frac12 \Delta_0 f_1 \,, \\
	K_1 &= \Delta_0 F_1 + 2 F_1 \,, &W_1 &= -r \partial_u L_1 + i\partial_z \Sigma_1 \,, &N_1 &= \frac14 \Delta_0\Delta_0f_1 + \frac12\Delta_0f_1 \,,\\
	H_1 &= \frac12 K_1 - r \partial_u F_1 \,,
\end{align}
while the quadratic perturbations read
\begin{align}
	L_2 &= i \partial_z f_2 + g \,, \\
	g &= \ax{\kappa} \partial_z(f_1^2) \,,\\
	P_2 &= P_0\Big(F_2 + \frac12 F_1^2 \Big) \,, \\
	\Sigma_2 &= \frac12 \Delta_0 f_2 + F_1 \Delta_0 f_1 + P_0^2 \Im(\partial_{\bar{z}} g + \partial_z f_1 \partial_u\partial_{\bar{z}} f_1) \,, \\
	K_2 &= \Delta_0 F_2 + 2 F_2 + 2F_1\Delta_0 F_1 + 2F_1^2 - 4P_0^2 \Im(\partial_{\bar{z}}f_1 \partial_u\partial_z F_1) + 2 P_0^2 \Re(\partial_z f_1 \partial_u^2\partial_{\bar{z}}f_1 - \partial_u \partial_z \bar{g}) \,, \\
	W_2 &= -r\partial_u L_2 + i \partial_z \Sigma_2 -i \partial_u(L_1\Sigma_1) \,, \\
	N_2 &= \!\begin{multlined}[t]\frac14 \Delta_0\Delta_0f_2 + \frac12\Delta_0f_2 + 2 F_1 \Delta_0 f_1 + \frac12 F_1 \Delta_0\Delta_0 f_1 + \frac12 \Delta_0(F_1\Delta_0 f_1) + \frac12 \Delta_0 f_1 \Delta_0 F_1 \\ + P_0^2 \Im(\partial_{\bar{z}} g + \partial_z f_1 \partial_u\partial_{\bar{z}} f_1) - \frac12 \Im\big[\Delta_0(P_0^2(\partial_{\bar{z}} g + \partial_z f_1 \partial_u\partial_{\bar{z}} f_1))\big] \\ - \frac12 P_0^2 \partial_u \Im\big[2 \partial_{\bar{z}} f_1 \partial_z \Delta_0 f_1 + \Delta_0 f_1 \partial_z\partial_{\bar{z}} f_1\big]\,, \end{multlined}\\
	H_2 &= \frac12 K_2 - r \partial_u F_2 + \frac{M \Sigma_1^2}{r^3} - \frac{N_1\Sigma_1}{r^2} \,.
\end{align}
Injecting expressions for the exact solutions we have obtained, given by Eq~\eqref{eq:sol-f1-F1} and Eq~\eqref{eq:sol-f2} and~\eqref{eq:sol-F2}, one obtains
\begin{align}
	L_1 &= i \ax{E} e^{\ax{\kappa} u} \partial_z Y_{\ax{\ell}}^{\ax{m}} \,, & P_1 &= \po{E} P_0 e^{-\po{\kappa} u} Y_{\po{\ell}}^{\po{m}}  \,, \nonumber\\
	\Sigma_1 &= - (\ax{\lambda} + 1) \ax{E} e^{\ax{\kappa} u} Y_{\ax{\ell}}^{\ax{m}} \,, &K_1 &= - 2 \po{\lambda} E_p e^{-\po{\kappa} u} Y_{\po{\ell}}^{\po{m}} \,, \nonumber\\
	W_1 &= -i(\ax{\kappa}r + \ax{\lambda} + 1) \ax{E} e^{\ax{\kappa} u} \partial_z Y_{\ax{\ell}}^{\ax{m}} \,, &N_1 &= 3M \ax{\kappa} \ax{E} e^{\ax{\kappa}u} Y_{\ax{\ell}}^{\ax{m}}\,,\nonumber\\
	H_1 &= (r \po{\kappa} -\po{\lambda}) E_p e^{-\po{\kappa} u} Y_{\po{\ell}}^{\po{m}} 
	\label{eq:metric-funcs-1st-order}
\end{align}
for the first-order contributions and
{
\allowdisplaybreaks
\begin{align}
	L_2 &= \sum_{\ell,m} \Big[ i \ax{E}\po{E} e^{(\ax{\kappa} - \po{\kappa})u} a_m^\ell + \ax{\kappa} \ax{E}^2 e^{2\ax{\kappa} u} k_{\ax{\ell}\ax{\ell}m}^{\ax{m}\ax{m}\ell} \Big] \partial_z Y_\ell^m \,,\nonumber\\
	P_2 &= P_0 \sum_{\ell,m} \Big[\po{E}^2 e^{-2\po{\kappa}u} \big(p_m^\ell + \frac12 k_{\po{\ell}\po{\ell}m}^{\po{m}\po{m}\ell}\big) + \po{E}\ax{E} e^{(\ax{\kappa}-\po{\kappa})u} r_m^\ell + \ax{E}^2 e^{2\ax{\kappa}u} q_m^\ell\Big] Y_\ell^m \,, \nonumber\\
	\Sigma_2 &= \sum_{\ell,m} \ax{E}\po{E} e^{(\ax{\kappa} - \po{\kappa})u} \big[-(\lambda+1)a_m^\ell - 2(\ax{\lambda} + 1) k_{\ax{\ell}\po{\ell}m}^{\ax{m}\po{m}\ell}\big] Y_\ell^m\,, \nonumber\\
	K_2 &=\!\begin{multlined}[t]\sum_{\ell,m} \Big[\po{E}^2 e^{-2\po{\kappa}u} \big\{ -2\lambda p_m^\ell -2(2\po{\lambda} + 1) k_{\po{\ell}\po{\ell}m}^{\po{m}\po{m}\ell} \big\} + \po{E}\ax{E} e^{(\ax{\kappa}-\po{\kappa})u} \big\{ -2\lambda r_m^\ell + 4\po{\kappa} \Im(h_{\po{\ell} \ax{\ell} m}^{\po{m} \ax{m} \ell})\big\} \\ + \ax{E}^2 e^{2\ax{\kappa}u} \big\{ -2\lambda q_m^\ell + 2 \ax{\kappa}^2 (\ell(\ell+1) k_{\ax{\ell}\ax{\ell}m}^{\ax{m}\ax{m}\ell} + h_{\ax{\ell} \ax{\ell} m}^{\ax{m} \ax{m} \ell}) \big\} \Big] Y_\ell^m \,,\end{multlined}\nonumber \\
	W_2 &=\!\begin{multlined}[t]\sum_{\ell,m} \Big[- i \ax{E}\po{E} e^{(\ax{\kappa} - \po{\kappa})u} \big\{r (\ax{\kappa} - \po{\kappa}) a_m^\ell + (\lambda+1)a_m^\ell +2(\ax{\lambda} + 1) k_{\ax{\ell}\po{\ell}m}^{\ax{m}\po{m}\ell} \big\} \\ +\ax{E}^2 e^{2\ax{\kappa} u} \big\{ - 2 \ax{\kappa}^2 r - \ax{\kappa}(\ax{\lambda} + 1) \big\} k_{\ax{\ell}\ax{\ell}m}^{\ax{m}\ax{m}\ell} \Big] \partial_z Y_\ell^m \,,\end{multlined} \nonumber\\
	N_2 &= \sum_{\ell,m} \ax{E}\po{E} e^{(\ax{\kappa} - \po{\kappa})u} \big[ 3M\kappa a_m^\ell + 2(\ax{\lambda} +1)(1 + \lambda + \ax{\lambda} + \po{\lambda}) k_{\ax{\ell}\po{\ell}m}^{\ax{m}\po{m}\ell} \big] Y_\ell^m\,,\nonumber\\
	H_2 &= \!\begin{multlined}[t]\sum_{\ell,m} \Big[\po{E}^2 e^{-2\po{\kappa}u} \big\{(2r\po{\kappa} -\lambda) p_m^\ell -(2\po{\lambda} + 1) k_{\po{\ell}\po{\ell}m}^{\po{m}\po{m}\ell} \big\} \\+ \po{E}\ax{E} e^{(\ax{\kappa}-\po{\kappa})u} \big\{ (-r(\ax{\kappa} - \po{\kappa})-\lambda) r_m^\ell + 2\po{\kappa} \Im(h_{\po{\ell} \ax{\ell} m}^{\po{m} \ax{m} \ell})\big\} \\ + \ax{E}^2 e^{2\ax{\kappa}u} \big\{ (-2r\ax{\kappa}-\lambda) q_m^\ell + \ax{\kappa}^2 (\ell(\ell+1) k_{\ax{\ell}\ax{\ell}m}^{\ax{m}\ax{m}\ell} + h_{\ax{\ell} \ax{\ell} m}^{\ax{m} \ax{m} \ell}) + \frac{\ax{\lambda} + 1}{r^3} M (3r\ax{\kappa} + \ax{\lambda} + 1) k_{\ax{\ell}\ax{\ell}m}^{\ax{m}\ax{m}\ell}\big\} \Big] Y_\ell^m\end{multlined}
	\label{eq:metric-funcs-2nd-order}
\end{align}
}
for the second-order contributions. This provides the explicit expressions for the different contributions to the perturbative radiative geometry.

\subsection{Covariant perturbation formalism}
\label{app:perts-BH}

Perturbations around a Schwarzschild BH can be decomposed into 10 functions $B_1$, ..., $B_{10}$, using the covariant form of perturbations described in~\cite{Thompson:2016fxe}. For a metric whose line element is
\begin{equation}
	\dd{s}^2 = \dd{s}_\mathrm{Schwa}^2 + h_{\mu\nu} \dd{x}^\mu \dd{x}^\nu \,,
\end{equation}
the perturbations tensor $h_{\mu\nu}$ is decomposed as
\begin{multline}
	h_{\mu\nu} = B_1 k_\mu k_\nu Y_\ell^m + 2 r B_2 k_{(\mu} \nabla_{\nu)} Y_\ell^m + 2 r B_3 k_{(\mu} \varepsilon\indices{^\rho_{\nu)}^\sigma^\lambda}k_\rho n^\sigma \nabla_\lambda Y_\ell^m + 2 B_4 k_{(\mu} n_{\nu)} Y_\ell^m + B_5 \varsigma_{\mu\nu} Y_\ell^m \\
	+ r^2 B_6 \big(\varsigma\indices{_\mu^\rho} \varsigma\indices{_\nu^\sigma} - \frac12 \varsigma_{\mu\nu} \varsigma^{\rho\sigma}\big) \nabla_\rho\nabla_\sigma Y_\ell^m + r^2 B_7 \varsigma\indices{_{(\mu}^\rho}\varepsilon\indices{^\sigma_{\nu)}_\lambda^\kappa} k_\sigma n^\lambda \nabla_\rho\nabla_\kappa Y_\ell^m + 2 r B_8 n_{(\mu} \nabla_{\nu)} Y_\ell^m \\
	+ 2 r B_9 n_{(\mu} \varepsilon\indices{^\rho_{\nu)}_\sigma^\lambda} k_\rho n^\sigma \nabla_\lambda Y_\ell^m + B_{10} n_\mu n_\nu Y_\ell^m \,,
	\label{eq:perts-covar}
\end{multline}
where the forms $k_\mu$, $n_\mu$ and $\varsigma_{\mu\nu}$ are defined such that
\begin{equation}
	k_\mu \dd{x}^\mu = - \dd{u} - \frac{\dd{r}}{A} \,,\quad n_\mu \dd{x}^\mu = \dd{r} \,,\quad \varsigma_{\mu\nu} \dd{x}^\mu \dd{x}^\nu = r^2 \dd{\Omega}^2 \,.
\end{equation}
Here, all functions $B_1$, ..., $B_{10}$ depend on $u$, $r$, and the integers $\ell$ and $m$. We do not write this explicitly in order to lighten the notations.

The functions $B_1$, ..., $B_{10}$ can be related to the functions used by Regge and Wheeler in their original work on Schwarzschild perturbations~\cite{Regge:1957td}. In this work, axial perturbations are represented by $h_0$, $h_1$ and $h_2$ while polar perturbations are described by\footnote{In the original work~\cite{Regge:1957td}, the notations $h_0$ and $h_1$ are used twice. Here, we keep $h_0$ and $h_1$ for the description of axial perturbations and we rename them respectively $\beta$ and $\alpha$ for the description of even perturbations.} $H_0$, $H_1$, $H_2$, $\alpha$, $\beta$, $K$ and $G$. The relations between these functions and the ones appearing in the covariant description~\eqref{eq:perts-covar} are
\begin{align}
	B_1 &= A H_0 \,, &B_2 &= -\frac{1}{r} \beta \,, & B_3 &= -\frac{1}{r} h_0 \,, & B_4 &= -H_1 \,, \nonumber\\
	B_5 &= K - \frac12 \ell(\ell+1) G \,, & B_6 &= G \,, &B_7 &= - \frac{1}{r^2} h_2 \,, & B_8 &= \frac{1}{r} \alpha \,, \nonumber\\
	B_9 &= \frac{1}{r} h_1 \,, & B_{10} &= A^{-1} H_2 \,. 
	\label{eq:link-Bi-RW}
\end{align}

In the $(u, r, z, \bar{z})$ coordinate system, using~\eqref{eq:perts-covar}, one obtains
\begin{multline}
	h_{\mu\nu} \dd{x}^\mu \dd{x}^\nu = B_1 Y_\ell^m \dd{u}^2 + \big(B_{10} - \frac{2}{A} B_4 + \frac{1}{A^2} B_1\big) Y_\ell^m \dd{r}^2 - 2\big(B_4 - \frac{B_1}{A}\big)Y_\ell^m\dd{u}\dd{r} \\
	-2 r \dd{u} \Re\big[(B_2 - i B_3) \partial_z Y_\ell^m \dd{z} \big] + 2 r \dd{r} \Re\Big[ \big((B_8 - \frac{B_2}{A}) -i (B_9 - \frac{B_3}{A}) \big)\partial_z Y_\ell^m \dd{z} \Big] \\
	+ r^2 B_5 Y_\ell^m \dd{\Omega^2} + 2 r^2 \Re\big[ (B_6 - 2 i B_7) (\partial_z^2 + 2 \partial_z(\log P_0) \partial_z) Y_\ell^m \dd{z}^2\big] \,.
	\label{eq:link-Bi-metric-perts}
\end{multline}

\subsection{Reconstructing Regge-Wheeler perturbations}
\label{sec:metric-reconstruction}

The computation of $f_1$, $f_2$, $F_1$ and $F_2$ allows us to write the line element~\eqref{eq:metric} perturbatively as
\begin{equation}
	\dd{s}^2 = \dd{s}_\mathrm{Schwa}^2 + \varepsilon h^{(1)}_{\mu\nu} \dd{x}^\mu \dd{x}^\nu + \varepsilon^2 h^{(2)}_{\mu\nu} \dd{x}^\mu \dd{x}^\nu + \mathcal{O}(\varepsilon^3) \,,
\end{equation}
with
\begin{equation}
	\dd{s}_\mathrm{Schwa}^2 = -A(r) \dd{u}^2 - 2 \dd{r}\dd{u} + r^2 \dd{\Omega}^2 \,,\quad A(r) = 1 - \frac{2M}{r} \,.
\end{equation}
We have introduced the first and second-order perturbations to a background Schwarzschild metric, respectively $h^{(1)}_{\mu\nu}$ and $h^{(2)}_{\mu\nu}$. These two contributions can be computed using the following expansions for the functions appearing in~\eqref{eq:metric} and presented in app~\ref{app:exp-metric}. 
%
%
The linear contribution decomposes as
\begin{align}
	h^{(1)}_{\mu\nu} \dd{x}^\mu \dd{x}^\nu & = -2 H_1 \dd{u}^2 + \frac{r}{2}\Big[  \Im(\Sigma_1) - 4 F_1 r\Big]\dd{\Omega}^2 \nn \\
	& \;\;\;\;\;- 4 \dd{r} \Re(L_1 \dd{z}) - 4 \dd{u} \Re\Big[\big(W_1 + L_1 A(r)\big)\dd{z}\Big] 
	\label{eq:h1-perts}
\end{align}
while the second order contributions is given by
\begin{align}
	h^{(2)}_{\mu\nu} \dd{x}^\mu \dd{x}^\nu & = -2 H_2 \dd{u}^2 - 4 \dd{r} \Re(L_2 \dd{z}) - 4 \dd{u} \Re\Big[\big(W_2 + L_2 A(r) + 2 H_1 L_1\big)\dd{z}\Big] -4 A \Re(L_1 \dd{z})^2 \nn  \\
	& \;\;\;\;\;  - 8 \Re(L_1 \dd{z}) \Re(W_1 \dd{z}) + \Big[\frac12 r \Im(\Sigma_2) - 2 F_2 r^2 + 2 F_1^2 r^2 - F_1 r \Im(\Sigma_1) + \Sigma_1 \bar{\Sigma}_1\Big]\dd{\Omega}^2 \,.
	\label{eq:h2-perts}
\end{align}
One can then compute the explicit expressions for the Regge-Wheeler metric perturbation functions in the case of perturbations given in \eqref{eq:h1-perts} and~\eqref{eq:h2-perts} by comparing~\eqref{eq:metric-funcs-1st-order},~\eqref{eq:metric-funcs-2nd-order} and~\eqref{eq:link-Bi-metric-perts}, using the definitions~\eqref{eq:link-Bi-RW}.

At first order, one obtains
\begin{align}
	&{}^{(1)}H_{0} = {}^{(1)}H_{2} = - {}^{(1)}H_{1} = - \frac{2}{A} (r \po{\kappa} - \po{\lambda}) \po{E} e^{-\po{\kappa} u} \,, \\
	&{}^{(1)}K = -2 \po{E} e^{-\po{\kappa} u} \,,\quad {}^{(1)}\alpha = {}^{(1)}\beta = 0 \,,\quad {}^{(1)}G = 0
\end{align}
for the even perturbations and
\begin{align}
	{}^{(1)}h_{0} & = -2 (\ax{\kappa}r + \ax{\lambda} + 1 - A) \ax{E} e^{\ax{\kappa}u} \,,\\
	 {}^{(1)}h_{1} & = \frac{2}{A} (\ax{\kappa}r + \ax{\lambda} + 1) \ax{E} e^{\ax{\kappa}u} \,,\\
	  {}^{(1)}h_{2} & = 0 \,
\end{align}
for the axial perturbations. We have not written explicitly the $\ell$ and $m$ dependency of these functions since only the $(\ell = \ax{\ell}, m = \ax{m})$ functions and $(\ell = \po{\ell}, m = \po{m})$ are non-zero respectively in the axial and polar cases. Noticed that at first order, the solution satisfies the RWZ gauge since
\be
{}^{(1)}G = {}^{(1)}h_{2} =  {}^{(1)}\alpha = {}^{(1)}\beta = 0
\ee
However, at second order, one should keep in mind that there are several contributions from different values of $(\ell, m)$. One obtains
\begin{align}
	{}^{(2)}H_{0, \ell}^m &= \!\begin{multlined}[t] -{}^{(2)}H_{1, \ell}^m = {}^{(2)}H_{2, \ell}^m =  -\frac{2}{A} \po{E}^2 e^{-2\po{\kappa}u} \big[(2r\po{\kappa} -\lambda) p_m^\ell -(2\po{\lambda} + 1) k_{\po{\ell}\po{\ell}m}^{\po{m}\po{m}\ell} \big] \\ -\frac{2}{A} \po{E}\ax{E} e^{(\ax{\kappa}-\po{\kappa})u} \big[ (-r(\ax{\kappa} - \po{\kappa})-\lambda) r_m^\ell + 2\po{\kappa} \Im(h_{\po{\ell} \ax{\ell} m}^{\po{m} \ax{m} \ell})\big] \\ -\frac{2}{A} \ax{E}^2 e^{2\ax{\kappa}u} \big[ (-2r\ax{\kappa}-\lambda) q_m^\ell + \ax{\kappa}^2 (\ell(\ell+1) k_{\ax{\ell}\ax{\ell}m}^{\ax{m}\ax{m}\ell} + h_{\ax{\ell} \ax{\ell} m}^{\ax{m} \ax{m} \ell}) + \frac{\ax{\lambda} + 1}{r^3} M (3r\ax{\kappa} + \ax{\lambda} + 1) k_{\ax{\ell}\ax{\ell}m}^{\ax{m}\ax{m}\ell}\big] \,, \end{multlined} \\
	{}^{(2)}\beta_{\ell}^m &= 2 \ax{\kappa} \ax{E}^2 e^{2\ax{\kappa}u} (2\ax{\kappa}r + \ax{\lambda} + 1 - A) k_{\ax{\ell}\ax{\ell}m}^{\ax{m}\ax{m}\ell}\,,\\
	{}^{(2)}\alpha_{\ell}^m &= -2 \ax{\kappa} \ax{E}^2 e^{2\ax{\kappa}u} (2\ax{\kappa}r + \ax{\lambda} + 2 - A) k_{\ax{\ell}\ax{\ell}m}^{\ax{m}\ax{m}\ell}\,,\\
	{}^{(2)}K_{\ell}^m &= \!\begin{multlined}[t] - 2 \po{E}^2 e^{-2\po{\kappa}u} \big(p^\ell_m - k_{\po{\ell}\po{\ell}m}^{\po{m}\po{m}\ell}\big) - 2 r^\ell_m \po{E}\ax{E} e^{(\ax{\kappa}-\po{\kappa})u} \\ + \ax{E}^2 e^{2\ax{\kappa} u} \big[-2 q^\ell_m + \frac{(\ax{\lambda}+1)^2}{r^2} k_{\ax{\ell}\ax{\ell}m}^{\ax{m}\ax{m}\ell} - \frac{A - 2(\ax{\kappa} r + \ax{\lambda} + 1)}{r^2} \big(A h_{\ax{\ell}\ax{\ell}m}^{\ax{m}\ax{m}\ell} - (\lambda + 1) j_{\ax{\ell} m}^{\ax{m} \ell}\big)\big] \,,\end{multlined}\\
	{}^{(2)}G_{\ell}^m &=  - \frac{1}{r^2} \ax{E}^2 e^{2\ax{\kappa} u} j_{\ax{\ell} m}^{\ax{m} \ell} \big[-A + 2(\ax{\kappa} r + \ax{\lambda} + 1)\big]\,,\\
	{}^{(2)}h_{0, \ell}^m &= 2 \ax{E}\po{E} e^{(\ax{\kappa}-\po{\kappa})u} \big[\big(A - r(\ax{\kappa}-\po{\kappa}) - \lambda - 1\big)a^\ell_m - 2(\ax{\lambda} + 1) k_{\ax{\ell}\po{\ell}m}^{\ax{m}\po{m}\ell} + 2 (r\po{\kappa} - \po{\lambda}) g_{\po{\ell}\ax{\ell}m}^{\po{m}\ax{m}\ell}\big] \,,\\
	{}^{(2)}h_{1, \ell}^m &= -2 \ax{E}\po{E} e^{(\ax{\kappa}-\po{\kappa})u} \big[\big(A - r(\ax{\kappa}-\po{\kappa}) - \lambda - 2\big)a^\ell_m - 2(\ax{\lambda} + 1) k_{\ax{\ell}\po{\ell}m}^{\ax{m}\po{m}\ell} + 2 (r\po{\kappa} - \po{\lambda}) g_{\po{\ell}\ax{\ell}m}^{\po{m}\ax{m}\ell}\big] \,,\\
	{}^{(2)}h_{2, \ell}^m &= 0 \,.
\end{align}
Notice that the second order perturbations do not satisfy all the RWZ gauge conditions since we have only ${}^{(2)}h_{2, \ell}^m =  0$. Thus only the axial part satisfies this gauge.

\section{A concrete illustrative example for a quadratic ASP}

\label{examp}

In this appendix, we present the detailed derivation of one concrete example in order to illustrate the efficiency of our approach. We consider linear perturbations with $\ax{\ell} = 2$, $\ax{m} = 1$ and $\po{\ell} = 3$, $\po{m} = 2$ which implies that their decay time and the associated $\lambda$ are given respectively by  
\begin{equation}
	\ax{\kappa} = \frac{2}{M} \,,\quad \po{\kappa} = \frac{10}{M}\,,\quad \ax{\lambda} = 2 \qq{and} \po{\lambda} = 5 \,.
\end{equation}
The first task is to compute explicitly the Clebsch-Gordan coefficients entering in the sources. For the present case, the relevant ones are given by
\begin{align}
	k_{220}^{110} &= \frac{1}{2\sqrt{\pi}} \,,& k_{220}^{112} &= \frac{1}{14} \sqrt{\frac{5}{\pi}} \,,& k_{222}^{112} &= \frac{1}{14} \sqrt{\frac{15}{\pi}} \,,& k_{220}^{114} &= - \frac{2}{7\sqrt{\pi}} \,,& k_{222}^{114} &= \frac{1}{7} \sqrt{\frac{5}{\pi}} \,.\nn \\
	k_{231}^{121} &= \frac12 \sqrt{\frac{3}{7\pi}} \,, &k_{231}^{123} &= \frac{1}{2\sqrt{6\pi}} \,, &k_{233}^{123} &= \frac16 \sqrt{\frac{5}{2\pi}} \,, &k_{231}^{125} &= -\sqrt{\frac{5}{231\pi}} \,, &k_{233}^{125} &= \frac13 \sqrt{\frac{10}{11\pi}} \,.\nn \\
	k_{330}^{220} &= \frac{1}{2\sqrt{\pi}} \,, &k_{330}^{224} &= - \frac{7}{22\sqrt{\pi}} \,, &k_{334}^{224} &= \frac{1}{22} \sqrt{\frac{35}{\pi}} \,, &k_{330}^{226} &= \frac{5}{11\sqrt{13\pi}} \,, &k_{334}^{226} &= \frac{5}{11} \sqrt{\frac{7}{13\pi}} \,.
\end{align}
Proceeding to the decomposition of the different source terms using the explicit expressions~\eqref{eq:expr-Alm},~\eqref{eq:expr-Plm} and~\eqref{eq:expr-Qlm}, one obtains for the polar-polar and axial-axial sources
\begin{align}
	\mathcal{A}^1_1 &= -96\sqrt{\frac{3}{7\pi}} \,, & \mathcal{A}^3_1 &= -26 \sqrt{\frac{6}{\pi}} \,, & \mathcal{A}^3_3 &= - 26 \sqrt{\frac{10}{\pi}} \,, & \mathcal{A}^5_1 &= 16 \sqrt{\frac{165}{7\pi}} \,,  & \mathcal{A}^5_3 &= - 16 \sqrt{\frac{110}{\pi}} \,,\nn \\
	\mathcal{P}^0_0 &= -\frac{120}{\sqrt{\pi}} \,, & \mathcal{P}^4_0 &= \frac{2380}{11\sqrt{\pi}} \,, & \mathcal{P}^4_4 &= -\frac{340}{11} \sqrt{\frac{35}{\pi}} \,, & \mathcal{P}^6_0 &= - \frac{5820}{11\sqrt{13\pi}} \,,  & \mathcal{P}^6_4 &= - \frac{5820}{11} \sqrt{\frac{7}{13\pi}} \,,\nn \\
	M^2 \mathcal{Q}^0_0 &= \frac{144}{\sqrt{\pi}} \,, & M^2\mathcal{Q}^2_0 &= \frac{384}{7} \sqrt{\frac{5}{\pi}} \,, & M^2\mathcal{Q}^2_2 &= \frac{384}{7} \sqrt{\frac{15}{\pi}} \,, & M^2\mathcal{Q}^4_0 &= - \frac{7136}{7\sqrt{\pi}} \,,  & M^2\mathcal{Q}^4_2 &= \frac{3568}{7} \sqrt{\frac{5}{\pi}} \,,
\end{align}
while for the coefficients $\mathcal{R}^\ell_m$ describing the polar-axial source, one finds
\begin{equation}
	M \mathcal{R}^4_{-3} = 100 \sqrt{\frac{10}{\pi}} \,,\qquad M \mathcal{R}^4_{-1} = 100 \sqrt{\frac{70}{\pi}}  \,.
\end{equation}
Gathering these different pieces, one can now evaluate explicitly the profile of the quadratic ASP $(F_2, f_2)$. They are given in the present case by
\begin{equation}
	f_2 = \po{E}\ax{E} \Big(-\sqrt{\frac{3}{7\pi}} Y_1^1 - \frac{13}{18\sqrt{6\pi}} Y_3^1 - \frac{13}{54}\sqrt{\frac{5}{2\pi}} Y_3^3 + \frac{2}{39}\sqrt{\frac{55}{21\pi}} Y_5^1 - \frac{2}{117}\sqrt{\frac{110}{\pi}} Y_5^3\Big) e^{-8u/M} \,.
\end{equation}
and
\begin{align}
	F_2 & = \po{E}^2 \Big(\frac{1}{2\sqrt{\pi}} Y_0^0 + \frac{119}{66\sqrt{\pi}} Y_4^0 - \frac{17}{66}\sqrt{\frac{35}{\pi}} Y_4^4 - \frac{97}{264\sqrt{13\pi}} Y_6^0 - \frac{97}{264}\sqrt{\frac{7}{13\pi}} Y_6^4\Big) e^{-20u/M} \\
	&  + \frac{\ax{E}^2}{M^2} \Big(\frac{3}{\sqrt{\pi}} Y_0^0 + \frac{16}{21}\sqrt{\frac{5}{\pi}} Y_2^0 + \frac{16}{7} \sqrt{\frac{5}{3\pi}} Y_2^2 - \frac{892}{357\sqrt{\pi}} Y_4^0 + \frac{446}{357}\sqrt{\frac{5}{\pi}} Y_4^2\Big) e^{4u/M} \\ 
	&+ \frac{\po{E}\ax{E}}{M} \Big(\frac{25}{33} \sqrt{\frac{5}{2\pi}} Y_4^{-3} - \frac{25}{33} \sqrt{\frac{35}{2\pi}} Y_4^{-1}\Big) e^{-8u/M} \,.
\end{align}

\end{document}